\documentclass[seceq,preprint]{ptptex}

\usepackage{graphicx}
\usepackage{wrapft}

\newcommand{\D}{\partial}

\newcommand{\F}{\frac}

\newcommand{\SR}{\!\!\!{/}}
\newcommand{\SRC}{\!\!\!\!{/}}
\newcommand{\be}{\begin{equation}}
\newcommand{\ee}{\end{equation}}
\newcommand{\bea}{\begin{eqnarray}}
\newcommand{\eea}{\end{eqnarray}}

\newcommand{\LP}[1]{\frac{\overrightarrow{\D}}{\D #1}}
\newcommand{\RP}[1]{\frac{\overleftarrow{\D}}{\D #1}}

\newcommand{\LQ}{\Lambda_{\mbox{\tiny QCD}}}
\newcommand{\LA}{\Lambda_{\mbox{\tiny AUX}}}

\newcommand{\qqquad}{\quad\qquad}
\newcommand{\qqqquad}{\qqquad\qqquad}

\newcommand{\Bq}{{\bf q}}
\newcommand{\Bp}{{\bf p}}

\newcommand{\Bx}{{\bf x}}

\newcommand{\TR}{^{\mbox{\tiny T}}}

\title{%        %You can use \\ for explicit line-break
The Chiral Phase Transition of the hot QCD \\
from the Exact Renormalization Group%
}

\author{%       %Use \scshape  for the family name
Dai \textsc{Shimizube}$^{1,}$\footnote{
E-mail address: simizube@phys.h.kyoto-u.ac.jp}
 and Jun-Ichi. \textsc{Sumi}$^{2,}$
\footnote{%
E-mail address: j-sumi@blue.vecceed.ne.jp}%
}

\inst{%         %Affiliation, neglected when [addenda] or [errata]
$^{1}$Graduate School of Human and Environmental Studies, Kyoto University \\
Yoshida, Kyoto University, Kyoto 606-8501, Japan \\
$^{2}$Riverroad Takano 221, Ooharada-cho 8, Ichijoji, Sakyo-ku,\\
Kyoto 606-8187, Japan
}

\abst{%         %this abstract is neglected when [addenda] or [errata]
The chiral phase transition of the finite temperature QCD is discussed in the ladder approximated 
Exact (Wilson's) Renormalization Group (ERG). The gluon thermal mass is incorporated by the Hard Thermal Loop (HTL) approximation. 
We calculated a critical temperature $T_c$, a critical exponent $\nu$, and order parameters such as chiral condensate $<\bar{\psi}\psi>$ and pion decay constants with this approximation.
The dependences of the pion decay constants on an unphysical parameter are fairly improved by including contributions from momentum dependent vertices. 
The critical temperature was found as: $T_c\approx 124$ MeV for $3$-flavors. 
It was shown that the critical temperature considerably depends on the approximation scheme for the gluon thermal
mass term. We also calculated the temperature dependence of 
the chiral condensate, $\pi$ decay constants, and the pion velocity.   
}

\begin{document}

\maketitle
\section{Introduction}
\label{sec:1}

Explorations of the chiral nature of QCD at high temperature are experimental as well as theoretical
interests. They are relevant to the early universe and to the heavy ion collisions at RHIC and 
those planned at LHC.
It is believed that the approximate chiral symmetry observed at bare Lagrangian dynamically
breaks down at zero temperature due to the chiral condensate contained in the QCD vacuum. 
At high temperature, the chiral condensate is expected 
to melt, and the chiral symmetry is restored at a certain critical temperature. 
Since the perturbative methods can not be applied to the non-trivial
QCD vacuum, so we need non-perturbative methods, such as the lattice Monte-Carlo
simulation, the Schwinger-Dyson equation, and the Exact Renormalization
Group, for investigating the vacuum structure of QCD at high temperature.  
    
For the zero temperature case, the Schwinger-Dyson equations (SDEs) has been studied 
\cite{rf:SDmain,rf:SDfourfermi,rf:SDreview,rf:SBtext}
and has been applied to the strong coupled QED and
QCD.\cite{rf:SDimproved1,rf:SDimproved2,rf:BSamp} \ Commonly,
the so-called (improved) ladder approximation \cite{rf:SDimproved1} has been applied to solve the SDEs. 
Although these approximation schemes seem to be fairly crude, the SDEs offer surprisingly good
results on the dynamical chiral symmetry breaking in QCD.
For the finite temperature gauge theories, the SDEs were also
applied.\cite{rf:SDEfiniteT1,rf:SDEfiniteT5,rf:SDEfiniteT6,rf:SDEfiniteT2,rf:SDEfiniteT3,rf:Takagi,rf:SDEfiniteT4} \ 
At finite temperature, the gauge interaction is corrected by thermal effects.
Thermal effects, such as the Debye screening and the Landau damping, appear in the gauge 
boson propagator as their thermal mass. 
So it is no longer to say that how to approximate the thermal mass is one of the main
subjects for the improvement of finite temperature analyses.
The hot QED was discussed in the instantaneous-exchange (IE) approximation 
for the gauge boson propagator including the screening
effect in the static limit,\cite{rf:SDEfiniteT1} \ or  the IE approximation for the longitudinal (electric) mode and 
the Hard Thermal Loop Approximation for transverse (magnetic) modes. \cite{rf:SDEfiniteT5} \ 
The chiral phase transition of the hot QCD 
was studied by neglecting the gluon thermal/Debye mass
term.\cite{rf:SDEfiniteT6,rf:SDEfiniteT2,rf:SDEfiniteT3} \ 
The Phase structure of hot and/or dense QCD was explored by introducing
the Debye mass in static limit of the gluon propagator.\cite{rf:Takagi}
\  
However unfortunately the improvement of the above approximations seems to be 
difficult due to the complexity of the integral equation.

The Exact Renormalization Group (ERG) has been applied to investigate the chiral nature of the strong
coupling gauge theory at zero temperature.\cite{rf:kanazawa} \  With ERG method, 
we can study the chiral critical behavior in very simple manner, and if we approximate the ERG flow
equations to the ladder-like corrections then it leads the identical results with those obtained
by solving the ladder SDEs. It should be noted that we can improve the approximation scheme systematically,
e.g. incorporating the ``crossed-ladder'' diagrams. Especially, the large gauge dependence of the solutions
obtained by the ladder SDEs\cite{rf:AoKugoBan} has been improved in the case of the chiral critical behavior 
in QED.\cite{rf:qed}  \ 

For the finite temperature, 
the chiral nature has been investigated by the ``Quark-Meson model'' in
ERG method.\cite{rf:ERGquark-meson} \ The Quark-Meson model is defined at `so-called' compositeness scale 
$k_{\rm\Phi}\approx600$ MeV from the zero temperature QCD and therefore does not incorporate 
the thermal screening of the gluon interaction. It is expected that the mesonic bound states
forms at the short scale region: $k>k_{\rm \Phi}$ and are not affected by the long distance: $k<k_{\rm \Phi}$ 
behavior of the gluon propagator. 
It is one of the issue whether the thermal screening effect to the chiral behavior 
can be neglected or not.

In this paper, we investigate the chiral phase transition of the hot QCD in the ladder approximation 
with including the gluon thermal mass in the Hard Thermal Loop (HTL) approximation. 
To introduce auxiliary fields, we consider a intermediate momentum scale similar to $k_{\Phi}$ in the Quark-Meson model. The pion decay constants depend on the intermediate scale, which is an unphysical parameter. By including contributions from momentum dependent vertices, the decay constants become considerably stable for the intermediate scale. 
The formation of the chiral condensate is 
remarkably affected by the thermal screening. 
Especially, the critical temperature is remarkably lowered in comparison with that given by 
neglecting the thermal mass. 
We also calculated the temperature dependence of the order parameters,
i.e. the chiral condensate and the pion decay constants for several approximation schemes for the thermal mass. 

We estimated the temperature dependence of the pion velocity too, from the ratio of the spatial
 component of the decay constant and the temporal component of that, which
 become different at finite temperature.\cite{rf:coolpion} \ 
 By the ladder approximated SDE, the temperature dependence of the `space-time averaged' 
component of the pion decay constant was
 evaluated,\cite{rf:SDEfiniteT2} \ however, that of the pion velocity was not.
It is expected that pions travel at less than the speed of light in a thermal bath.
In this paper, it is shown that pions are slowed down slightly near $T_c$. \  
 
 This paper is organized as follows. We explain our approximation scheme and 
derive the ERG flow equations in \S\ref{sec:2}. 
In \S\ref{sec:3} we show the numerical results. 
\S\ref{sec:4} is devoted to the summary and discussions.

\section{The ERG equation}
\label{sec:2}
In this section we explain the approximation scheme first, and next we derive the ERG equation. 
Let us start form $N_f$-flavor QCD at finite temperature $T$. 
The bare Lagrangian density ${\cal L}$ is given by
\be
\label{eq:bareL}
{\cal L}=\F14{\bf tr}F_{\mu\nu}F^{\mu\nu}+\bar qi D\SRC \,q 
+\F1{2\alpha}\left(\partial_\mu A^A_\mu\right)^2
+\partial_\mu\bar C^A\left[\partial_\mu\delta_{AB}+gf_{ABC}A^C_\mu \right]C_B,
\ee
where $q$ denotes the $N_f\times N_c$ components quarks. 
$N_c$ is the number of color.
The field strength $F_{\mu\nu}=F^A_{\mu\nu}T^A$ and 
the covariant derivative $D_\mu$ are given by,
\bea
&&D_\mu=\partial_\mu+igA^A_\mu T^A,\\
&&F^A_{\mu\nu}=\partial_\mu A^A_\nu-\partial_\nu A^A_\mu+gf_{ABC}A^C_\mu A^B_\nu.
\eea
The third term is gauge fixing one, and the fourth is from FP ghosts.
We choose the Landau gauge $\alpha=0$. Here ${\cal L}$ is invariant under
the chiral flavor symmetry $U_R(N_f)
\times U_L(N_f)$\footnote{The axial chiral symmetry $U(1)_A$ is broken due to the anomaly, 
and the QCD phase transition is affected by the instanton effect.
However, we do not consider the instanton effect in this paper.
} . 
The axial current is defined as $J_{5\mu}^a\equiv\bar q\gamma_\mu\gamma_5\lambda^aq$, where 
$a$ denotes iso-spin indices and $\lambda^a$ are the generators of the
chiral $U(N_f)$ (${\bf tr}\lambda^a\lambda^b=\F12\delta^{ab}$). 

\subsection{Gauge interaction}
There are some problems in the application of the ERG to the gauge theory
because of the infrared cutoff term which explicitly
breaks gauge invariance. \cite{rf:gaugeERG,rf:gaugekeep} \ 
There are two ways to treat the gauge theory. One is generalizing the
ERG keeping the gauge symmetry.\cite{rf:gaugekeep} \ This is of course 
an ideal method, however their formulations are not accompanied with the non-perturbative
approximation method. The other is introducing the gauge non-invariant
counter terms to compensate the gauge invariance. Hence, the bare action does
not satisfy the Slavnov-Taylor Identity, but Modified one. \cite{rf:gaugeERG} \  
However, since we would like to introduce the gluon self energy in the HTL approximation, 
we must calculate the momentum dependent gluon self energy and therefore 
relevant terms of the Modified Slavnov-Taylor Identity (MSTI)  become
complicated integral equations.

\begin{figure}[htb]
\centerline{\includegraphics[width=7.5cm]{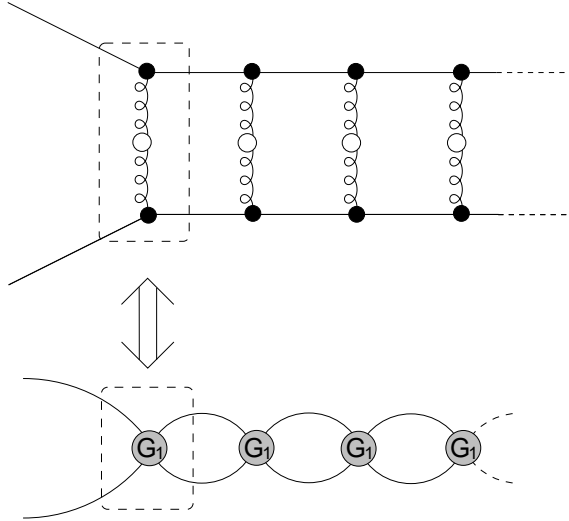}}
%\epsfxsize=0.50\textwidth
%\centerline{\epsfbox{ladder2.eps}}
\caption{1-gluon exchange as 4-fermi interaction in the ladder approximation.}
\label{fig:ladder}
\end{figure}

In this paper, we do not adopt the either methods explained above. 
We integrate out gluons at the beginning and solve ERG flow equations for the reduced fermionic theory. 
This procedure generally causes many multi-fermi interactions. 
By the diagrammatical consideration, one can easily realize that all ladder type diagrams are generated 
by only the four fermi operator induced by the one gluon exchange. 
Other multi fermi operators generate the non-ladder diagrams. 
So as shown in Fig. \ref{fig:ladder}, we replace the gauge sector 
with the non-local four fermi interaction induced by the one gluon
exchange i.e.
\be
G_1\equiv\int_q\int_p\int_kg(p,k)g(q,-q-p-k)
\bar q(k)\gamma_\mu T^Aq(p)
D_{\mu\nu}^{AB}(p+k)
\bar q(q)\gamma_\nu T^Bq(-q-p-k),
\label{eq:onegluon}
\ee
where $g(p,k)$ and $D_{\mu\nu}^{AB}$ are the running gauge coupling constant and 
the gluon propagator respectively. Note that, 
since other corrections from the gluon fluctuations do not contribute to the ladder approximation,
our treatment does not mean the additional approximation. 
Now, what we should investigate is reduced to the fermionic theory interacting through 
the non-local four fermi interaction
(\ref{eq:onegluon}) and the ladder approximate solutions can be found in the leading order of
$1/N_c$ expansion with the coupling constant $\bar g$: $g^2\equiv \bar g^2/N_c$.

The flow of the momentum dependent four fermi interaction is studied by Meggiolaro et al.
\cite{rf:Meggiolaro} \ 
They do not integrate out gluons at the beginning. They introduce 
the IR cutoff term for gluons and the ERG flow starts from the
QCD classical action. 
In their work, to compensate the gauge invariance, 
the gluon mass is given so as to satisfy the MSTI
at the leading perturbative order.  
We do not examine the flow of such interactions themselves as they did, but the flow of the
effective potential $V$. The contribution from the momentum dependent four fermi interaction appears 
in the ERG equation for $V$ and affects the chiral transition, in which we are interested.

The argument of the gauge coupling constant should be 
taken as $g(p,k)=g((p+k)^2)$ to satisfy the chiral Ward identities.\cite{rf:chiralWT} \ 
The infrared singularity of the gauge coupling constant is regularized as :
\be
g^2(\Lambda)=\left\{
\begin{array}{ll}
\F{1}{b\ln(\Lambda^2/\LQ^2)} & {\rm if}\;\Lambda>\Lambda_1\\
\F{1}{b\ln(\Lambda_1^2/\LQ^2)} 
+\F{\left((\ln(\Lambda_2/\Lambda))^2-(\ln(\Lambda_2/\Lambda_1))^2\right)}
{b\ln(\Lambda_2/\Lambda_1)(\ln(\Lambda_1^2/\LQ^2))^2}
 &  {\rm if}\;\Lambda_1\ge\Lambda>\Lambda_2\\
\F{1}{b\ln(\Lambda_1^2/\LQ^2)} 
-\F{\ln(\Lambda_2/\Lambda_1)}{b(\ln(\Lambda_1^2/\LQ^2))^2}
 &  {\rm if}\;\Lambda\le \Lambda_2
\end{array}
\right.
\ee
where $\ln(\Lambda_1/\LQ)=0.25$, 
$\ln(\Lambda_2/\LQ)=-1.0$ and $b=(11N_c-2N_f)/48\pi^2$. 

At high temperature gluons become massive by the thermal fluctuation.
We incorporate the Debye-screening effect to the gluon propagator
$D_{\mu\nu}^{AB}$ in the HTL approximation. Of course, the HTL approximation is based on the
perturbative QCD and it is valid in the high energy (short distance) region.
However, it is expectable that the short distance behavior is important to the QCD chiral
physics, not the long distance one. Also, it is known that the order parameters, such as 
the chiral condensate, 
hardly depend on the difference of the treatments of the running coupling constant in the
low energy region.\cite{rf:BSamp} \ 
So applying the HTL approximation for the gluon propagator is a proper procedure
for our present purpose
, which is the first step of  the improvement of the approximation for introducing thermal effects.
Of course it should be checked that how order parameters and the critical temperature depend on the IR cutoff scheme for $g(\Lambda)$
at finite $T$. We show the $\Lambda_1$ dependence of our numerical results later, in \S\ref{sec:3}.

The gluon self-energy is a gauge dependent quantity, contrary to the photon self-energy
in QED. However the HTL approximated one has no dependence on the choice of gauge.
At finite temperature, the self-energy matrix takes the following form: 
\be 
\Pi_{\mu\nu}^{AB}=\left(FP^L_{\mu\nu}+GP^T_{\mu\nu}\right)\delta^{AB}.
\ee
The projectors $P^L_{\mu\nu}$ and $P^T_{\mu\nu}$ are given by 
\be 
P^T_{\mu\nu}=\bar\delta_{\mu\nu} -\F{\bar p_\mu\bar p_\nu}{\bar p^2},\quad 
P^L_{\mu\nu}=\delta_{\mu\nu} -\F{p_\mu p_\nu}{ p^2}-P^T_{\mu\nu}.
\ee 
We use the following notation, $\bar p_\mu=(0,\vec p)$ and $\bar\delta_{\mu\nu}={\rm diag}(0,1,1,1)$. 
One can easily verify that $P^T_{\mu\mu}=2,P^L_{\mu\mu}=1$.
In the HTL approximation, $F$ and $G$ take the form:\cite{rf:LeBellac} \ 
\bea
&&F=p^2+\F{p^2}{\bar p^2}\Pi_{00}(p_0,\bar p)
=p^2+m_{\mbox{\tiny HTL}}^2(\eta^2-1)\Phi(\eta),\\
&&G=p^2+\F{1}{2}\Pi_{ii}(p_0,\bar p)
-\F{p_0^2}{2\bar p^2}\Pi_{00}(p_0,\bar p)
=p^2+\F12m_{\mbox{\tiny HTL}}^2\left[1-(\eta^2-1)\Phi(\eta)\right],
\eea
where $\eta=ip_0/|\bar p|$.  At temperature $T$, the thermal mass $m_{\mbox{\tiny HTL}}$ and $\Phi(\eta)$ are given by
\bea
&&m_{\mbox{\tiny HTL}}^2=\F13g^2T^2\left(
N_c+\F12N_f\right)
,\\
&&\Phi(\eta)=\F\eta 2\ln\left(\F{\eta+1}{\eta-1}
\right)-1\nonumber\\
&&\qquad\;=\F{p_0}{|\bar p|}\tan^{-1}\F{|\bar p|}{p_0}-1.
\eea

\subsection{The auxiliary fields}

It is convenient to introduce the auxiliary fields for the
quark-antiquark pairs to make calculation efficient.\cite{rf:kanazawa} \ 
We introduce the auxiliary fields for the composite operators 
$\bar q_{\small{\LA}}q_{\LA}$ and $i\bar q_{\LA}\gamma_5q_{\LA}$, 
where $q_{\LA}$ is the low energy mode of the quarks defined by
\be
q_{\LA}(x)=T\sum_n\int \F{ d^3\Bp}{(2\pi)^3}q_i(\Bp)e^{ipx}\theta (\LA-p ) .
\ee
The $q_0$ integral reduces to the Matsubara sum, i.e. $q_0\rightarrow 2(n+1)\pi T, n=0,\pm 1,\pm 2,\cdots$.
The intermediate momentum scale $\LA$ plays similar role to the compositeness scale $k_{\rm \Phi}$
in Ref.\citen{rf:ERGquark-meson}. However we do not impose the
compositeness conditions \cite{rf:BHL} at this scale. 

The auxiliary fields  
$\sigma$ and $\pi$ are introduced by inserting the following trivial Gaussian integrals: 
\bea
\label{eq:GaussSigma}
&&1={\cal N}^{-1}\int D\sigma\exp\left\{ -\F12M^2
\int^{\beta}_0d\tau\int d^3\Bx \left(\sigma-yM^{-2}\bar q_{\LA}q_{\LA}\right)^2\right\},\\
\label{eq:GaussPi}
&&1={\cal N}^{-1}\int D\pi\exp\left\{ -\F12M^2
\int^{\beta}_0d\tau\int d^3\Bx \left(\pi-yM^{-2}i\bar q_{\LA}\gamma_5q_{\LA}\right)^2\right\},
\eea
where ${\cal N}$ is a irrelevant constant and $\beta\equiv1/kT$~\footnote{In the following, we choose units such that 
Boltzmann constant $k=1$.}. Since the compositeness conditions are not imposed, 
$M^2$ and $y$ are the arbitrary constants. 
Of course, the physical quantities should be independent of $M^2$, $y$
and $\LA$. 
We introduced the composite operators whose tensor structures are ${\bf 1}$ 
and $\gamma_5$, and
neglected the others, like the Pagels-Stokar approximation.\cite{rf:PagelsStokar} \ 
Using $G_1$ of (\ref{eq:onegluon}) and the contribution from
(\ref{eq:GaussSigma}) and (\ref{eq:GaussPi}), we can write our Lagrangian density
${\cal L}_A$ as
\be
{\cal L}_A=\bar qi\D\SR q + V_a
+G_1,\label{eq:bareL2}
\ee 
where
\bea
&&V_a=\F12M^{-2}y^2 \left( (\bar q_{\small{\LA}}q_{\LA})^2+(i\bar
q_{\LA}\gamma_5q_{\LA})^2\right)\nonumber\\
\label{eq:Vaux}
&&\qqquad-y(\sigma \bar q_{\small{\LA}}q_{\LA} +\pi i\bar
q_{\LA}\gamma_5q_{\LA})+\F12 M^2(\sigma^2+\pi^2).
\eea
The ERG evolution is divided into two regions, 
$\Lambda>\LA$ and $\Lambda<\LA$. In  the high energy region
$\Lambda>\LA$, ${\cal L}_A$ describes a pure fermionic system interacting through $G_1$. 
At $\Lambda=\LA$, additional Yukawa interactions and four fermi interactions are switched on. 
In $\Lambda<\LA$, our model becomes the quark-meson system with $G_1$. 
We separate quark fields into the high energy mode and  the low energy mode at the scale
$\LA$, and the auxiliary fields for mesons are introduced as what is coupled with the low energy mode, 
$q_{\LA}$. We show the $\LA$ dependence of the order parameters in \S\ref{sec:3}.

The intermediate scale $\LA$ plays almost similar role for $k_{\rm \Phi}$ in the Quark-Meson model,
\cite{rf:ERGquark-meson} \ however $\LA$ is not an UV cutoff for some effective theory. 
We solve the ERG flow equations from the sufficiently large cutoff $\Lambda_0\gg\LA$ 
for each temperature $T$. In both regions: $p>\LA$ and $p<\LA$, 
the gluonic contributions are explicitly evaluated through one gluon exchange term $G_1$, 
since we are interested in the influence of the gluon thermal screening effect for 
the chiral phase transition.

\subsection{Flow equations in the ladder approximation}

Now, let us derive the ERG equation. The flow equation describes the
change of the effective average action $\Gamma_\Lambda$ under an
infinitesimal shift of the infrared momentum cutoff $\Lambda$.
We start with the construction of $\Gamma_\Lambda$.
The generating functional of the connected Green function is
\be
W_{\Lambda}[J]=\ln\int D\bar{q}Dq\exp\{ -S_{\rm cut}-S_{\rm bare}+\int_0^\beta d\tau\int
d^3 {\bf x}(\bar\eta q-\bar q\eta)\},
\label{eq:connect}
\ee
where $S_{\rm bare}$ is the bare action, i.e. $S_{\rm bare}=\int d\tau d^3{\bf x}
{\cal L}_A$, and $\{ \bar\eta , \eta \}$ are external sources. 
 The IR cutoff term $S_{\rm cut}$ is introduced for
the (anti-)quark fields,
\be
S_{\rm cut}[\bar q,q]
=\int_0^\beta d\tau\int d^3{\bf x}~
\bar q\Delta^{-1}(-i\D,\Lambda)q.
\label{eq:cut}
\ee
Here $\Delta^{-1}$ is a chiral invariant cutoff operator and has the property,
\be
\Delta^{-1}(p,\Lambda)
%=C^{-1}(p/\Lambda)p\SR
\longrightarrow\left\{
\begin{array}{cc}
0 & {\rm for}\quad p\gg\Lambda\\
\infty & {\rm for}\quad p\ll\Lambda.\\
\end{array}
\right.
\ee 
We do not introduce the IR cutoff term for boson fields in this paper since we treat the
ladder part only. In the ladder approximation, the bosonic degrees of freedom, 
$\sigma$ and $\pi$ are not integrated in (\ref{eq:connect}).
So we can forget the infrared problem for mesons at $T<T_c$. 
The infrared problem originated from the absence of the
IR cutoff for gluons can be regularized by introducing a sufficiently
small IR cutoff $\Lambda_{\rm IR}$ to the gluon propagator in $G_1$.
However, our results are not affected by $\Lambda_{\rm IR}$, since our ERG evolutions
of the effective potential as well as the order parameters are frozen
fast when $\Lambda$ gets smaller than the typical scale for the chiral physics.

The cutoff effective action $\Gamma_{\Lambda}$ is defined by Legendre transformation
\be
\label{eq:Legendre}
\Gamma_{\Lambda}[\Phi]=-W_{\Lambda }[J]+\int_0^\beta d\tau\int d^3{\bf x}
J\cdot\Phi +S_{\rm cut}[\Phi],
\ee
where $\Phi$ is given by $\delta W_{\Lambda }[J]/ \delta J$ and we use the shorthand notation 
of the external sources: $J=\{ \bar\eta, \eta\TR \}$.
Note that, since $S_{\rm cut}$ preserves 
the chiral symmetry, the effective 
action also respects it. 

Differentiating both sides of Eq.(\ref{eq:connect}) and using (\ref{eq:Legendre}), 
we get the ERG equation for the effective average action,\cite{rf:EvolutionEq} \ 
\be
\Lambda\frac{d}{d\Lambda}\Gamma_\Lambda[ \Phi]
=-\frac{1}{2}{\bf Str}\left[\Lambda\frac{d}{d\Lambda}\Delta^{-1}\left(
\Delta^{-1}+
\Gamma_\Lambda^{(2)}\right)^{-1}\right],\label{eq:smoothRG}
\ee
where ${\bf Str}$ is super-trace which involves momentum (or coordinate)
integration, Matsubara summation, spinor summation and color summation. 
$\Gamma_\Lambda^{(2)}$ is a second (functional) derivative 
with respect to 
the fields $\Phi\TR=(q\TR,\bar q)$, i.e.
\be
\left(\Gamma_\Lambda^{(2)}\right)_{xy}\equiv
\frac{\overrightarrow\delta}{\delta\Phi\TR_{x}}
\Gamma_\Lambda[\Phi]
\frac{\overleftarrow\delta}{\delta\Phi_{y}}\;.
\ee

We employ the sharp cutoff scheme in this paper since in the smooth cutoff case we must perform the
momentum integration with respect to the spatial momenta numerically for each Matsubara mode.
We want to avoid such a tiresome and unpractical work. The sharp cutoff limit of 
Eq.~(\ref{eq:smoothRG}) can be written as:\cite{rf:MomentumExp} \ 
\be
\label{eq:ERG1PI}
\Lambda\F{d}{d\Lambda}\Gamma_{\Lambda}=-\F\Lambda2{\bf Str}\left[
\F{\delta(|q|-\Lambda)}{\gamma(\Bq)}\widehat\Gamma^{(2)}\left(
1+C\widehat\Gamma^{(2)}
\right)^{-1}
\right],
\ee
where $\widehat\Gamma^{(2)}$ is a interaction part of the second derivative
of $\Gamma_{\Lambda}$ and $\gamma(\Bq)$ is an inverse propagator:
\be
\gamma(\Bq)\equiv
\left(\begin{array}{cc}
   0 &  q\SR \\
 q\SR\TR & 0
\end{array}
\right)\; .\label{eq:IP}
\ee
The infrared cutoff propagator $C(\Bq)$ is given by,
\be
C(\Bq)=\F{\theta(|q|-\Lambda)}{\gamma(\Bq)}.
\ee

It is known that, in the sharp cutoff case the vertices of the effective action 
$\Gamma_\Lambda$ should be expanded in powers of
$|p|\equiv\sqrt{p^2}$ instead of the derivatives $\D_\mu$.\cite{rf:MomentumExp} \ Heaviside $\theta$ function in ERG equations is also expanded in terms of $|p|$ as
\be
\theta(|p-q|-\Lambda)=\theta(p^2+2p\cdot q)=\theta(\hat p\cdot q)
+\sum\F{p^n}{n!}\delta^{(n)}(\hat p\cdot q),
\ee
where $\hat p$ is a unit vector parallel to $p$.
Integrating r.h.s. of the ERG equation(\ref{eq:ERG1PI}) with respect to the internal momenta $q$, one can expand it
in terms of the momentum scale $|p|$.

We approximate the effective action as:
\be
\label{eq:defK}
\Gamma_\Lambda=\int_0^\beta d\tau\int d^3x
\left\{
\bar qi\D\SR q 
+V(\bar q,q,\sigma,\pi)+V_a+G_1+K(\pi,i\bar{q}\gamma_5q)\right\},
\ee
where $K(\pi,i\bar{q}\gamma_5q)$ is the higher order term of momentum scale expansion. The explicit form of $K$ is shown later, in (\ref{eq:KP}). 
We neglect the kinetic term of $\sigma$, which is not concerned with the analysis here. 
The potential part $V(\bar q,q,\sigma,\pi)$ is a function not only of 
$S=\bar qq$ and $P=i\bar q\gamma_5q$ but also of 
$\bar q_{\LA}q_{\LA}$ and $i\bar q_{\LA}\gamma_5q_{\LA}$, therefore vertices of $V$ have momentum
dependence like $\theta (\LA -p)$.

The complexity arising from $\theta (\LA -p)$ can be avoided by dividing the ERG evolution into two regions, i.e.
the high energy region $\Lambda>\LA$ and the low energy region $\Lambda<\LA$.
In the high energy region, $\bar q_{\LA}q_{\LA}$ and $i\bar q_{\LA}\gamma_5q_{\LA}$
do not contribute to r.h.s. of Eq.(\ref{eq:ERG1PI}) since the supports of $\bar q_{\LA}(p)$ 
and $q_{\LA}(p)$ are restricted to $p<\LA<\Lambda$. Thus we can forget $V_a$ in this region.
Next, in the low energy region, we need not distinguish $q_{\LA}(p)$ from $q(p)$ in the vertices 
of the effective action with external momentums $p_i<\LA$ and of course in the ERG equation
(\ref{eq:ERG1PI}). Hence we can replace $q_{\LA}(p)$ with $q(p)$ by choosing $\LA$
larger than typical scales of the chiral dynamics.
Now our prescription is that in the high energy region we solve ERG equation by 
neglecting $V_a$ term and at $\Lambda=\LA$ we shift the potential 
$V(\bar q,q,\sigma,\pi)\to V(\bar q,q,\sigma,\pi)+V_a(\bar q_{\LA}=\bar q,q_{\LA}=q)$. 
Hereafter in both regions, we write the arguments of the potential $V$ as $S$, $P$,
$\sigma$ and $\pi$ i.e. $V(S,P,\sigma,\pi)$

We can get the ERG equation for $V$ by 
substituting the zero modes of the composite operators $S_0\equiv\int d\tau d^3{\bf x}S(x)$ 
and $P_0\equiv\int d\tau d^3{\bf x}P(x)$\footnote{
For the fermionic theory, we can not insert the zero modes of the fermion fields.
Otherwise, the higher vertices which is necessary to produce the bulk structure of the general 
Green function of the fermions vanish due to the Grassmann nature of fermions.}. 
The ERG equation for $V$ can be read,\cite{rf:MomentumExp} \ 
\be
\Lambda\F{d}{d\Lambda}V=-\F\Lambda2{\bf Str}\left[\ln\left(
\gamma(\Bq)+V^{(2)}+G_1^{(2)}
\right)\right],\label{eq:LPAERG}
\ee
where $V^{(2)}$ and $G_1^{(2)}$ are the matrices of 
the second derivative of $V$ and of the 
four fermi interaction induced by one gluon exchange (\ref{eq:onegluon})
with respect to the (anti-)quark fields respectively. 

The derivatives of $V$ with respect to (anti-)quarks can be written as 
\bea 
&&\LP{\bar q_{I,a}}V\RP{q^{J,b}}= \delta^a_b\delta^I_J
\F{\D}{\D S}V(S,P)+i\delta^a_b\delta^I_J\gamma_5
\F{\D}{\D P}V(S,P)
+\cdots,\label{qd1}\\
&&\LP{q^{\mbox{\tiny T} I,a}}V\RP{\bar q\TR_{J,b}}=-\delta^b_a\delta^J_I
\F{\D}{\D S}V(S,P)-i\delta^a_b\delta^I_J\gamma_5
\F{\D}{\D P}V(S,P)
+\cdots,\label{qd2}
\eea
where we omit the terms generating the non-ladder diagrams in ERG. 
Thus the matrix $V^{(2)}$ is 
\be
V^{(2)}=
\left(\begin{array}{cc}
0
& \delta^a_b\delta^I_J\left(V_S+i\gamma_5V_P\right) \\
-\delta^b_a\delta^J_I\left(V_S+i\gamma_5\TR V_P\right)  & 0
\end{array}
\right)\; ,\label{eq:V2}
\ee
where subscripts `$S$'  and `$P$'  denote the derivative 
with respect to $S\equiv\bar qq$ and $P\equiv i\bar q\gamma_5q$ respectively,
e.g. $\bar V_S=\D\bar V/\D S$.
The second functional derivative of the gauge induced four fermi interaction $G_1^{(2)}$ is calculated as:
\be
G_1^{(2)}=g^2\left(D^{-1}\right)_{AB}^{\mu\nu}
\left(\begin{array}{cc}
T^A\gamma_\nu q\otimes q\TR\gamma_\mu\TR {T^B}\TR
&   -T^A\gamma_\nu q\otimes \bar q\gamma_\mu T^B
 \\
-{T^A}\TR\gamma_\nu\TR\bar q\TR\otimes q\TR\gamma_\mu\TR {T^B}\TR &   
{T^A}\TR\gamma_\nu\TR\bar q\TR\otimes \bar q\gamma_\mu T^B
\end{array}
\right)\; .\label{eq:mgauge}
\ee
The argument of the gauge coupling constant $g$ is an I.R. cutoff $\Lambda$, 
i.e. leading order in the momentum scale expansion. 
The ladder contributions to $\sigma$ channel and $\pi$ channel are given
as follows, 
\be
-T^A\gamma_\nu q^{I}\otimes \bar q_{J}\gamma_\mu T^A
=-\F1{8N_c^2N_f}\delta_{\mu\nu}\delta_J^I\delta^{AA}
(\bar qq+i\gamma_5\bar q i\gamma_5q)~{\bf 1}_c+\cdots,
\ee
where we use the relation of the generator $T^A$:
\be
(T^A)_i^j(T^A)_k^l=\F12\delta_k^j\delta_i^l-\F1{2N_c}\delta_i^j\delta_k^l,
\ee
and ${\bf 1}_c$ is a color unit matrix.
Consequently, we find the ladder contribution from the gauge interaction $G_1$ as:
\be
G_1^{(2)}=-\F{g^2(\Lambda)}{8N_c^2N_f}\left(D^{-1}\right)_{AA}^{\mu\mu}
\left(\begin{array}{cc}
0
&   \delta_b^a\delta_J^I(S+i\gamma_5P)
 \\
-\delta_a^b\delta_I^J(S+i\gamma_5\TR P) &   
0
\end{array}
\right)\; .\label{eq:G2}
\ee

The trace of the gluon propagator ${\bf
tr}D^{-1}=(D^{-1})_{AA}^{\mu\mu}$ reduces to a factor
\be
{\bf tr}D^{-1}=\F{8}F+\F{16}G
=\F{8}{\Lambda^2+m_L^2(p_0/|\bar p|)}+\F{16}{\Lambda^2+m_T^2(p_0/|\bar p|)}
\ee
for $N_c=3$, where $m_L$ and $m_T$ are longitudinal component and
transverse component of thermal mass respectively. The argument $p_0/|\bar p|$ of thermal mass
is written in terms of $\Lambda$ and $\omega_n=(2n+1)\pi T$ : 
$(\sqrt{\Lambda^2/\omega_n^2-1})^{-1/2}$ in the ERG flow equation. 
The bosonic Matsubara frequency is $\omega_n=2n\pi T$. 
However, in our analysis, the momentum which flows into $G_1$ in (\ref{eq:onegluon}) is expected
to $(2n+1)\pi T$ because of the derivative expansion. We expand $G_1((2n+1)\pi T+\omega_l)$ in terms of
$\omega_l$, which is odd, and include only the leading part: $l=0$.
Inserting Eqs.~(\ref{eq:IP}), (\ref{eq:V2}) and (\ref{eq:G2}) to Eq.(\ref{eq:LPAERG}), 
the final expression of the potential part of the ERG equation reduces to 
\be
\label{eq:ERGpoFT}
\Lambda\F\partial{\partial\Lambda}V=
\F{\Lambda^2T}{\pi^2} N_fN_c{\sum_n}'\sqrt{\Lambda^2-\omega_n^2}
\ln\left(\Lambda^2+\bar V_S^2 + \bar V_P^2
\right),
\ee 
where we write $\bar V\equiv V-g^2(S^2+P^2){\bf tr}D^{-1}/32N_c^2N_f$ and 
${\sum_n}'$ is a summation over Mastubara modes $n$ with the condition $\Lambda^2\ge\omega_n^2$. 
Thus the RG evolution terminates at $\Lambda=\pi T$.

%%%%%%%%%%%%%%%%%%%%%%%%%%%%%%05/5/21%%%%%%%%%%%%%%%%%%%%%%%%%%%%%%%%%%%%%%%%%%%%
The pion wave function renormalization factors of the spatial derivative and the temporal one 
are different because the Lorentz invariance is lost. 
For the derivation of the ERG flow equations for the $\pi$ wave function
renormalization factors, $Z_{\pi L}$ and $Z_{\pi T}$, we consider the higher order contribution of momentum scale expansion in (\ref{eq:defK}),
\be
\label{eq:KP}
K(\pi,P)=
\F12Z_\pi(p)\pi_\Bp\pi_{-\Bp}
+Y_\pi(p)\pi_\Bp P_{-\Bp}
+\F12G(p)P_\Bp P_{-\Bp},
\ee
where $P_\Bp$ is a pionic composite operator: $P_\Bp=(\bar
qi\gamma_5q)_\Bp$, and subscripts $\Bp$ and $-\Bp$ denotes their
momenta. $Y_{\pi}(p)$ and $G(p)$ are the momentum dependent
vertices
of Yukawa interaction and that of four fermi interaction respectively.
Taking functional derivative of $K(\pi,P)$ with respect to
$P_{\Bp}$ and $P_{-\Bp}$, and substituting to (\ref{eq:ERG1PI}), the ERG
equation for $K$ is given by

\be
\label{eq:ERGgammaPI}
\Lambda\F{d}{d\Lambda}K(\Bp)=
\F\Lambda 2{\bf Str}\bigl[
\delta(|q|-\Lambda){\bf S}_F(\Bq)\hat{{\bf Y}}
\theta(|p+q|-\Lambda){\bf S}_F(\Bp+\Bq)\hat{{\bf Y}}\bigr]\F{\delta \hat{K}}{\delta P_{-\Bp}}\F{\delta \hat{K}}{P_\Bp}\theta(\LA-\Lambda).
\ee
$\hat{K}$ denotes the `interaction parts' of $K$, where the pion kinetic term is subtracted in (\ref{eq:KP}). 
 The step function $\theta(\LA-\Lambda)$ in r.h.s. of Eq.(\ref{eq:ERGgammaPI}) is due
to the cutoff of the quark-meson interaction. 
It tells that $K$ contributes to the whole ERG flow for $\Lambda < \LA$. 
${\bf S}_F$ and $\hat{{\bf Y}}$ are the massive quark propagator on the vacuum 
$<\sigma >=\sigma_0$ and the matrix part of the Yukawa coupling constant:
$Y_{\pi}(p)\equiv Y(p)\hat{{\bf Y}}$ respectively. Introducing the coupling
constants $V=m~\bar qq+
%y~\pi\bar qi\gamma_5q+
\cdots$, ${\bf S}_F$ is written as:
\be
{\bf S}_F^{-1}(\Bq)=\gamma(\Bq)+\left(\begin{array}{cc}
 {\bf 0}&  m\\
 -m
& {\bf 0}
\end{array}
\right)\; ,
\ee
and $\hat{{\bf Y}}$ is given by,
\be
\hat{{\bf Y}}\equiv \left(\begin{array}{cc}
 {\bf 0}&  i\gamma_5\\
 -i\gamma_5\TR
& {\bf 0}
\end{array}
\right)\; .
\ee
To devive the ERG equation of $Z_{\pi L}$ and $Z_{\pi T}$, we expand
(\ref{eq:ERGgammaPI}) in powers of momentum scale
to second order. Note that, terms which mix $p_0$ and $\bar{p}$ do not
contribute to the wave function renormalization factors.
The momentum dependent vertices are expanded as:
\bea
&&Z_{\pi}(p)=Z_{\pi L}p_0^2+Z_{\pi T}\bar{p}^2+O(p), \\
\label{eq:mseY}
&&Y(p)=(|p_0|/\Lambda)y_{1L}+(\bar{p}/\Lambda)y_{1T}+(p_0/\Lambda)^2y_{2L}+(\bar{p}/\Lambda)^2y_{2T},\\
\label{eq:mseG}
&&G(p)=(|p_0|/\Lambda)g_{1L}+(\bar{p}/\Lambda)g_{1T}+(p_0/\Lambda)^2g_{2L}+(\bar{p}/\Lambda)^2g_{2T}.
\eea
We need to be careful about expanding the inside of square bracket in
(\ref{eq:ERGgammaPI}) in terms of $p_0$, which takes
discontinuous value at finite temperature.
Let us expand a function $f(p_0)$ as:
\be
f(p_0)=f_0+f_1|p_0|+f_2p_0^2.
\ee
We estimate the expansion of $f(p_0)$ using three points of bosonic Matsubara frequency, $p_0=0$, $p_0=2\pi T$ 
and $p_0=4\pi T$, as:
\bea
\label{eq:disexpansion1}
&&4\pi Tf_1=-3f(p_0=0)+4f(p_0=2\pi T) - f(p_0=4\pi T) , \\
\label{eq:disexpansion2}
&&8\pi^2 T^2f_2=f(p_0=0)-2f(p_0=2\pi T)+f(p_0=4\pi T).
\eea
The result of the expansion above becomes identical to that of the ordinary
momentum scale expansion at $T\rightarrow 0$. 
Finally, we get following ERG equations,
\bea
\label{eq:ERGZPILT}
&&\Lambda\F{d}{d\Lambda}Z_{\pi(L,T)}
=m^2I_{2(L,T)}+(2my_{2(L,T)}+y_{1(L,T)}^2)I_0+2my_{1(L,T)}I_{1(L,T)},\\
\label{eq:ERGy1}
&&\Lambda\F{d}{d\Lambda}y_{1(L,T)}=mg_{\sigma}I_{1(L,T)}+(y_{1(L,T)}g_{\sigma}+mg_{1(L,T)})I_0,\\
&&\Lambda\F{d}{d\Lambda}y_{2(L,T)}=mg_{\sigma}I_{2(L,T)}+(y_{1(L,T)}g_{\sigma}+mg_{1(L,T)})I_{1(L,T)}\nonumber\\
\label{eq:ERGy2}
&&\qqqquad+(mg_{2(L,T)}+y_{1(L,T)}g_{1(L,T)}+y_{2(L,T)}g_{\sigma})I_0,\\
\label{eq:ERGg1}
&&\Lambda\F{d}{d\Lambda}g_{1(L,T)}
=g_{\sigma}^2I_{1(L,T)}+2g_{\sigma}g_{1(L,T)}I_0,\\
\label{eq:ERGg2}
&&\Lambda\F{d}{d\Lambda}g_{2(L,T)}
=g_{\sigma}^2I_{2(L,T)}+2g_{\sigma}g_{1(L,T)}I_{1(L,T)}+(2g_{\sigma}g_{2(L,T)}+g_{1(L,T)}^2)I_0.
\eea
The threshold functions  $I_{k(L,T)}$ include Matsubara sum. The explicit forms
of $I_{k(L,T)}$ are given in Appendix \ref{appendix:A}. 
ERG equations for $m$ and $g_{\sigma}$ are derived by the polynomial expansion of $V$ in terms of the composite operator. We show the explicit form later. 

\section{Numerical Analysis}
\label{sec:3}
\subsection{Truncation and $\LA$ dependence of the order parameters}
\label{sec:zero}
Before starting the analysis of the chiral phase transition at finite
temperature, we investigate the truncation and $\LA$ dependence of the order 
parameters. 
In general, the approximate solutions depend $\LA$ since the $\LA$ dependent factors can not 
be eliminated by the normalization of the redundant degrees of freedom $\sigma$, $\pi$. Hence 
the $\LA$ dependence of the results should be verified. 
In this paper we show the truncation/$\LA$ dependence of the order
parameters
such as the pion decay constant $f_\pi$ and the chiral condensate $<\bar \psi \psi>$ at zero temperature. 

At zero temperature, the ERG flow equation for $V$ reduces to 
\be
\Lambda\F\partial{\partial\Lambda}V=\F1{4\pi^2} N_fN_c
\ln\left(\Lambda^2+\bar V_S^2+\bar V_P^2
\right),
\ee
where $\bar V\equiv V-3g^2(S^2+P^2)/2N_c^2N_f\Lambda^2$. This flow equation leads to the 
ladder-exact results for the chiral condensate $<\bar\psi\psi>$
and the dynamical quark mass $m_{\rm eff}$.\cite{rf:kanazawa} \ Here $(\bar\psi)\;\psi$ denotes the 
(anti-)quark field with a definite flavor. 

To estimate the $\pi$ decay constant, 
we must calculate the wave function renormalization factor of the meson fields $Z_{\pi}$.
The ERG equations of $Z_{\pi}$ and momentum dependent vertices are the
same form as (\ref{eq:ERGZPILT})-(\ref{eq:ERGg2}). At zero temperature,
there exists no distinction between L and T. The explicit forms of zero
temperature coefficients $I_k$ are also given in Appendix \ref{appendix:A}.

The $\pi$ decay constant $f_\pi$ is given in terms of $Z_\pi$ and the vacuum expectation value (VEV) $\sigma_0$ \cite{rf:ERGquark-meson}
\footnote{We introduced flavor-singlet auxiliary fields by (\ref{eq:GaussSigma}) and (\ref{eq:GaussPi}). In our approximation, the decay constant can be regarded as that of meson flavor-$SU(3)$ octet.}
:
\be
f_\pi=\sqrt{2/3}Z_\pi^{1/2}\sigma_0.
\ee
The coefficient differs from that in \citen{rf:ERGquark-meson} because our normalizations of the auxiliary fields are different. 
To estimate the chiral condensate, we introduce the external source $m_0$ 
for the composite operator $\bar q q$. The chiral condensate is given by
\be
\label{eq:def_condensate}
-<(\bar\psi\psi)_{\Lambda_0}>=\F1{N_f}\F{\D V(\sigma_0)}{\D m_0}\Big|_{m_0=0},
\ee
where we write the renormalization point $\Lambda_0$ since this formula leads the chiral condensate 
renormalized at U.V. cutoff scale $\Lambda_0$. 
The ERG flow starts from $\Lambda=\Lambda_0$ , 
so the solutions of the ERG depend on $\Lambda_0$. 
In this paper, we set $\Lambda_0$ as: $\Lambda_0=\LQ\times e^{10.5}$.  Hereafter, we consider the case: $N_f=3$. 

\begin{figure}[htb]
\centerline{\includegraphics[width=9.75cm]{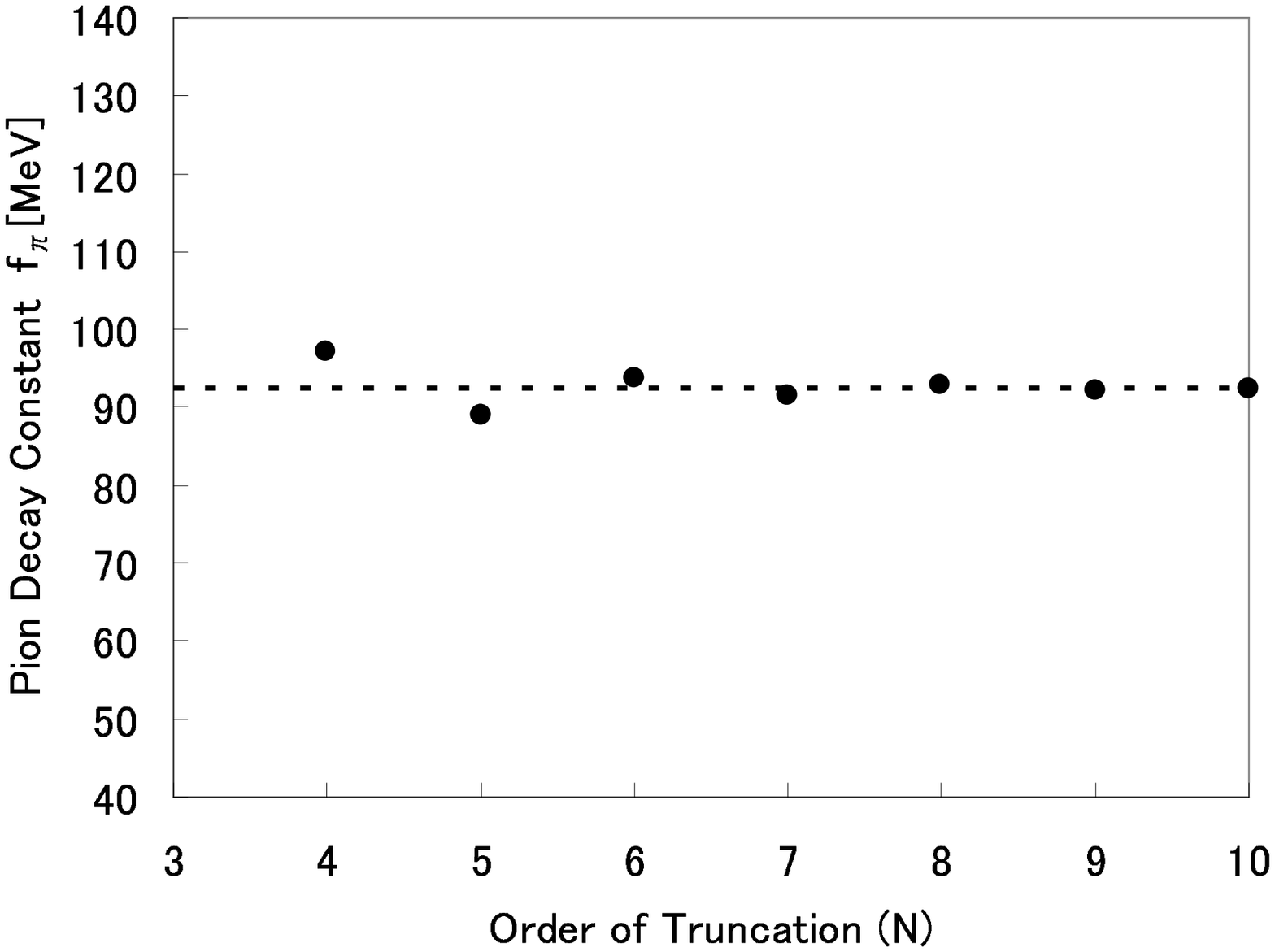}}
%\epsfxsize=0.65\textwidth
%\centerline{\epsfbox{fpi.eps}}
\caption{Truncation dependence of the $\pi$ decay constant. 
The dotted line is $f_\pi=92.42$ MeV. The order of truncation corresponds to $N$.}
\label{fig:1}
\end{figure}

We first investigate the truncation dependence of the order parameters.
The potential $V$ is truncated to the $(N-1)$-th polynomial of $S$,
\be
\label{eq:truncate}
V(S)=\sum_{n=1}^{N}v_{n-1}S^{n-1},
\ee
where $v_n$ are the VEV $\sigma_0$ dependent coefficients. By virtue of the parity invariance,
the higher terms of $P$ do not contribute to $O(P^0)$ terms. So we can
neglect such terms in (\ref{eq:truncate}) without losing generality.
 The ERG flow equation reduces to
a coupled ordinary differential equation. 
The truncation ($N$) dependence of $f_\pi$
is shown in Fig.\ref{fig:1}. 
As shown in Fig.\ref{fig:1}, $\pi$ decay constant rapidly 
converges against truncation of the effective potential $V(S)$. 
We set $\LQ$ to $583$ MeV, 
so as to the convergent $f_{\pi}$ becomes $92.42$ MeV,
which is the experimental value at $T=0$. \cite{rf:fpiex} \ 
Throughout this paper, $\LQ$ is fixed to this value. 
In this calculation, we choose $\Delta t\equiv\ln(\LA/\LQ)=1.0$.

%\vskip5mm

In Table \ref{table:1}, the truncation dependence of the
chiral condensate, the quark dynamical mass and the $\pi$ decay constant are described. 
The chiral condensate is renormalized at 1 GeV. 
Although the chiral condensate and the dynamical quark mass have already been
calculated in Ref.\citen{rf:kanazawa},
we show the truncation dependence of these quantities since the results may depend on the unphysical
parameters, e.g. $\Delta t$, for the lower truncation $N\le5$ where the results do not converge yet.
\begin{table}[htb]
\caption{Truncation dependence of the
chiral condensate, the quark dynamical mass and the $\pi$ decay constant. $\psi$ denotes the quark 
with a definite flavour.}
\label{table:1}
\begin{center}
\begin{tabular}{c|ccccccc}
\hline 
\hline 
Truncation ($N$)& 
$4$ & $5$ & $6$ & $7$ & $8$ & $9$ & $10$\\
\hline
$|<\bar\psi\psi>|^{1/3}$ (MeV)& 
$273.2$ & $258.2$ & $260.8$ & $259.6$ & $260.2$ & $259.9$ & $259.9$ \\
$m_{\rm eff}$ (MeV)&
$1072.4$ & $1084.8$ & $1063.4$ & $1073.1$ & $1068.6$ & $1070.6$ & $1069.8$ \\
$f_\pi$ (MeV)&
$97.24$ & $88.88$ & $93.70$ & $91.44$ & $92.72$ & $92.13$ & $92.42$ \\
\hline
\end{tabular}
\end{center}
\end{table}

\begin{figure}[htb]
\centerline{\includegraphics[width=9.15cm]{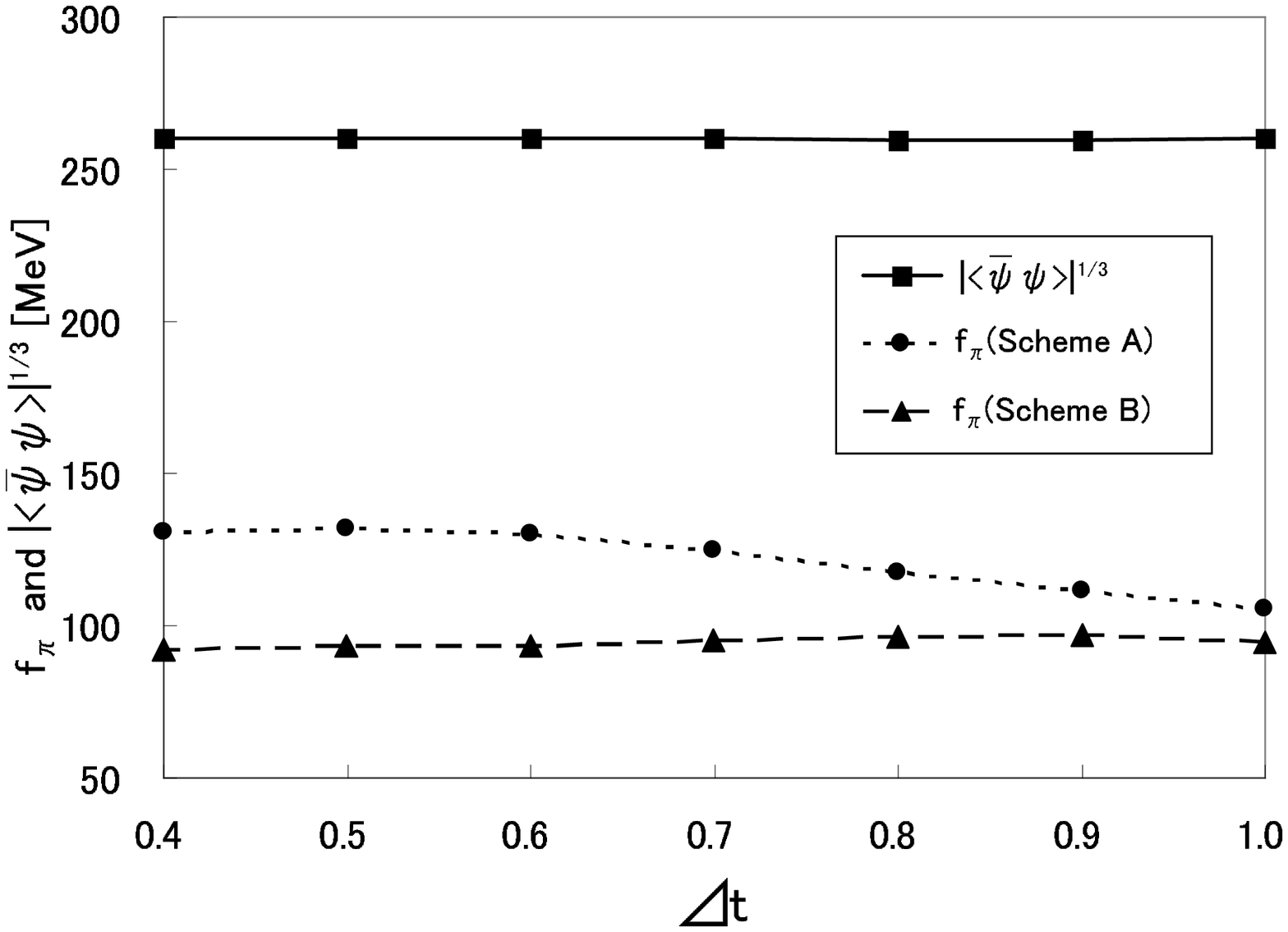}}
%\epsfxsize=0.61\textwidth
%\centerline{\epsfbox{fpit.eps}}
\caption{$\LA$ dependence of the $\pi$ decay constant and the chiral condensate for the truncation $N=10$.}
\label{fig:2}
\end{figure}

Next, we investigate the $\LA$ dependence of the order parameters.
The $\LA$ dependence of $<\bar\psi\psi>^{1/3}$ and $f_\pi$ for $N$=10 are
summarized in Table \ref{table:2} and 
Fig. \ref{fig:2}. 
The chiral condensate is quite stable to $\Delta t$.
%%%%%%%%%%%%%%%%%%%%%%05/5/23%%%%%%%%%%%%%%%%%%%%%%%%%%%%%%%%%%%%%%%%%%%%%%%
We show the $\LA$ dependence of the pion decay constants in two
approximation schemes. In `Scheme A' we neglect $Y(p)$ and $G(p)$ in
(\ref{eq:KP}).
`Scheme B' denotes the solution of (\ref{eq:ERGZPILT})-(\ref{eq:ERGg2}).
The $\LA$ dependence of $\pi$ decay constant is considerably decreased
by the momentum scale expansion, 20\% to 4\%.
%%%%%%%%%%%%%%%%%%%%%%%%%%%%%%%%%%%%%%%%%%%%%%%%%%%%%%%%%%%%%%%%%
 O($\D^0$) quantity $<\bar\psi \psi >^{1/3}$ and $m_{\rm eff}$ are stable to
the change of $\LA$ in this approximation.
The stabilities of these quantities are maintained at finite temperature.
For $T\approx 120$ MeV and $N=6$, $\LA$ dependence of $<\bar \psi \psi >^{1/3}$ was
estimated $0.5$\% in the region  $\Delta t=[0.5,1.0]$.
Therefore the critical temperature is also stable.
Hereafter we choose $\Delta t=1.0$ and $N=6$.

\begin{table}[htb]
\caption{$\LA$ dependence of the order parameters.}
\label{table:2}
\begin{center}
\begin{tabular}{c|ccccccc}
\hline 
\hline 
$\Delta t$ & $0.4$ & $0.5$ & $0.6$ & $0.7$ & $0.8$ & $0.9$ & $1.0$\\
\hline
$|<\bar\psi\psi>|^{1/3}$ (MeV) & $260.2$ & $260.1$ & $260.3$ & $260.3$
 & $259.9$ & $260.0$ & $260.2$\\
$f_\pi$[Scheme A] (MeV)& $130.47$& $131.94$& $130.17$& $124.67$& $117.59$
& $111.69$& $105.30$\\
$f_\pi$[Scheme B](MeV)& $94.49$&$96.65$&$96.35$&$95.08$&$93.50$&$93.21$&$92.42$\\
\hline
\end{tabular}
\end{center}
\end{table}

\subsection{Chiral phase transition at high temperature}

Let us analyze the chiral phase transition at high
temperature. 
We first solve the ERG equation for the potential $V(S)$ to estimate 
the critical temperature and the chiral condensate.
The ERG equation for $V(S)$ is already derived in (\ref{eq:ERGpoFT}).
Expanding the potential as: $V=v_0+mS-g_\sigma S^2/2+G_6S^3/3+\cdots$, and
substituting to the ERG equation we have
\bea
&&\Lambda\F d{d\Lambda}v_0=\zeta^0\ln(\Lambda^2+m^2),\\
\label{eq:v1}
&&\Lambda\F d{d\Lambda}m=-\F{2m\zeta^1}{\Lambda^2+m^2},\\
\label{eq:v2}
&&\Lambda\F d{d\Lambda}\left(-\F{g_\sigma}2\right)=
\F1{\Lambda^2+m^2}\left(\zeta^2+2mG_6\zeta^0
-\F{2m^2\zeta^2}{\Lambda^2+m^2}\right),\\
&&\Lambda\F d{d\Lambda}\left(\F{G_6}3\right)=-\F{2G_6\zeta^1}{\Lambda^2+m^2}
+\F{2m}{(\Lambda^2+m^2)^2}\left\{
\zeta^3+2mG_6\zeta^1
-\F83\F{m^3\zeta^3}{\Lambda^2+m^2}
\right\}
, \\
&&\Lambda\F d{d\Lambda}\left(\F{G_8}4\right)=\cdots,
\eea
where $\zeta^l$ denotes Matsubara frequency sums i.e.
\be
\zeta^l=\F{\Lambda^2T}{\pi^2}N_cN_f{\sum_n}'\sqrt{\Lambda^2-\omega_n^2}
\left(g_\sigma+g^2{\bf tr}D^{-1}/16N_c^2N_f\right)^l,
\ee
where $N_f=N_c=3$.

\begin{figure}[thb]
\centerline{\includegraphics[width=9.15cm]{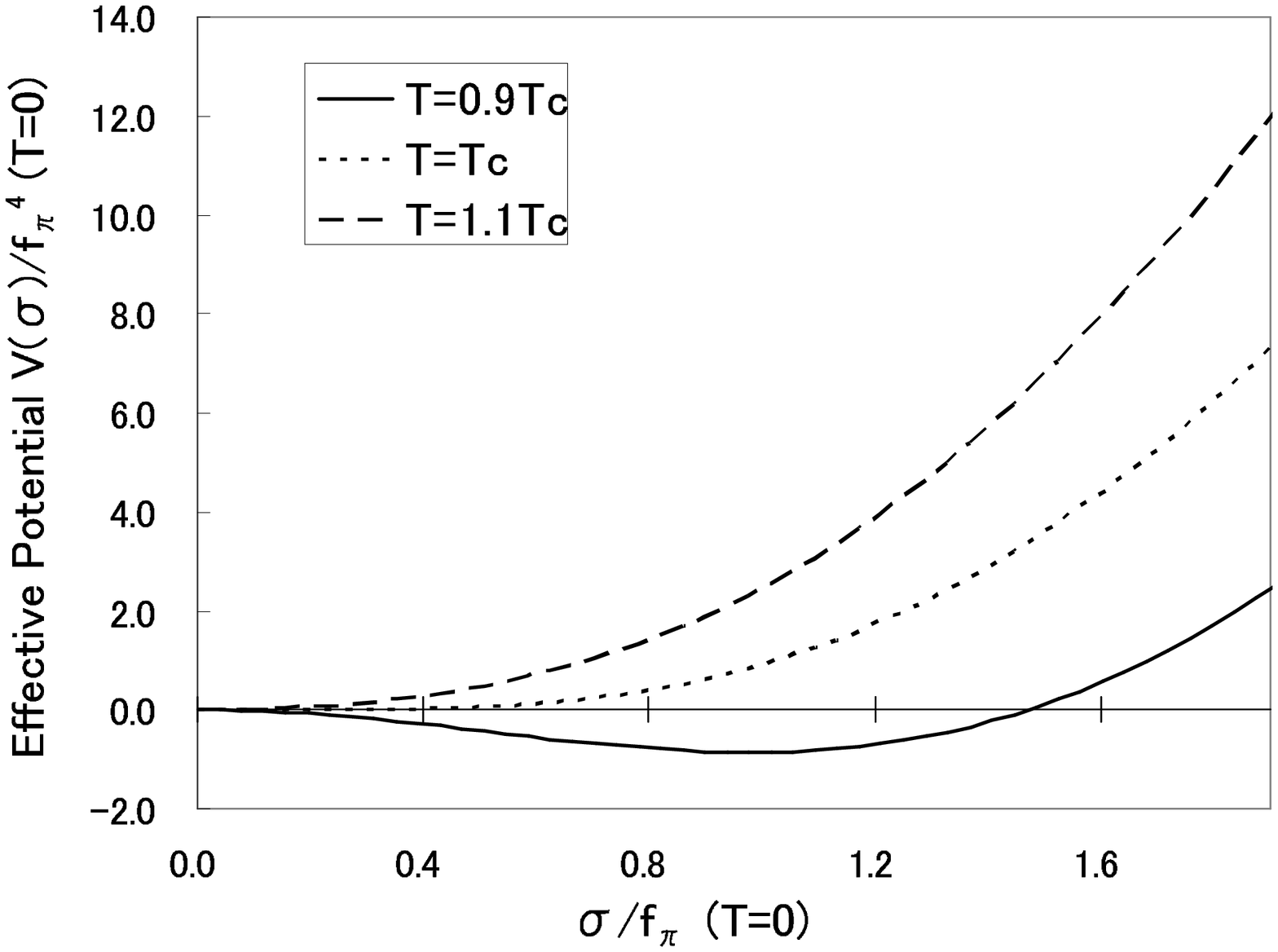}}
%\epsfxsize=0.61\textwidth
%\centerline{\epsfbox{t-depen-pot.eps}}
\caption{Temperature dependence of the effective potential. The gluon thermal mass is incorporated.}
\label{fig:3}
\end{figure}

The temperature dependence of the potential $V$ are shown in Fig.\ref{fig:3}.
The order of phase transition is second order or very weak first order, and
the critical temperature in our approximation is found to be $124.1$ MeV.  
This result should be compared with $T_c=169$ MeV given by Y. Taniguchi and 
Y. Yoshida \cite{rf:SDEfiniteT6}, $T_c=166$ MeV given by M. Harada  and A. Shibata\cite{rf:SDEfiniteT2} and/or
$T_c=129$ MeV proposed by O. Kiriyama, M. Maruyama and
F. Takagi.\cite{rf:SDEfiniteT3} \ Their analyses do not consider the
gluon thermal mass. 
In Ref.\citen{rf:Takagi}, the effect of the Debye screening at finite
temperature and/or density is studied. The thermal mass is included in
the static limit of the HTL approximation, thus the magnetic screening
is ignored. $T_c$ gets $147$ MeV by this approximation. \ 
These works adopt the same running coupling constant 
$g(p,k)$
at $T=0$. Furthermore, the angular dependent part of
the argument of $g(p,k)=g((p+k)^2)$ was neglected.
In our analysis, the higher order 
terms with respect to $|p|$ are not introduced at present.
We consider the leading part of derivative expansion, and the argument of $g$ becomes $\Lambda$ in our analysis.
At $T=0$, this leading approximation scheme is identical to the prescription 
 $g(p,k)=g(\max(p,k))$ in SDEs.\cite{rf:kanazawa}
   
Next, we estimate the pion decay constants at finite temperature.
At finite temperature, there are two distinct $\pi$ decay constants: the 
temporal one $f_{\pi L}$ and the spatial one $f_{\pi T}$.
They are defined by
\bea
&&j_5^0\sim f_{\pi L}\D^0\pi+\cdots,\\
&&j_5^i\sim f_{\pi T}\D^i\pi+\cdots.
\eea
We can calculate these in terms of the wave function renormalization factor of meson field\footnote{We renormalize the pion field as: $\pi_R=Z_{\pi L}^{1/2}\pi$.} :
\be
\label{eq:fpiFT}
f_{\pi L}=\sqrt{2/3}Z_{\pi L}^{1/2}\sigma_0,\qquad
f_{\pi T}=\sqrt{2/3}Z_{\pi T}/Z_{\pi L}^{1/2}\sigma_0.
\ee
The ERG flow equations for $Z_{\pi L}$ and $Z_{\pi T}$ are already
derived in (\ref{eq:ERGZPILT}). 
Solving those equations, and substituting the solutions to (\ref{eq:fpiFT}),
we gain $f_{\pi L}$ and $f_{\pi T}$.   

\begin{figure}[htb]
\centerline{\includegraphics[width=12.15cm]{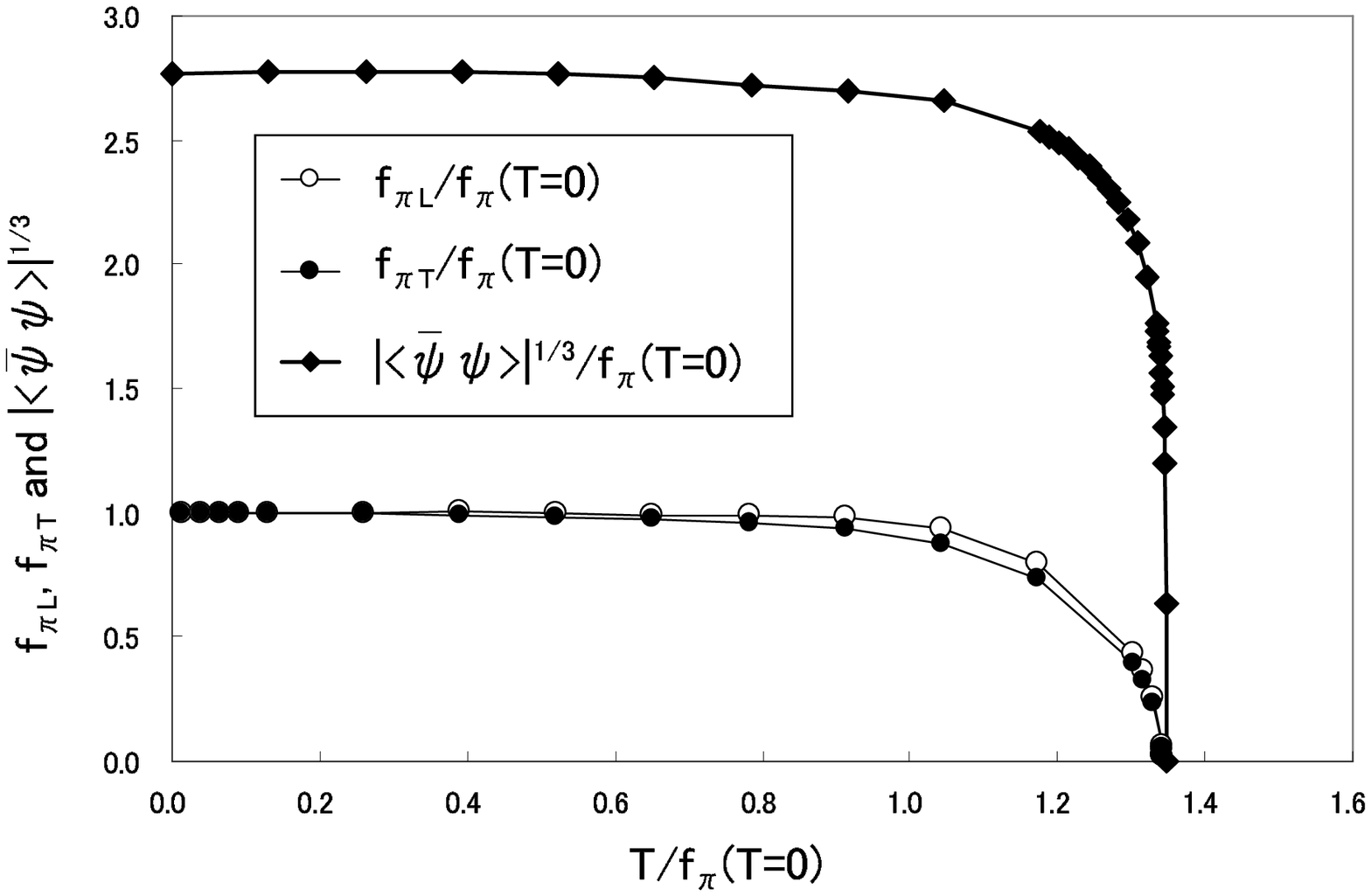}}
%\epsfxsize=0.81\textwidth
%\centerline{\epsfbox{t-depen.eps}}
\caption{Temperature dependence of the order parameters.}
\label{fig:4}
\end{figure}

\begin{figure}[hbt]
\centerline{\includegraphics[width=9.15cm]{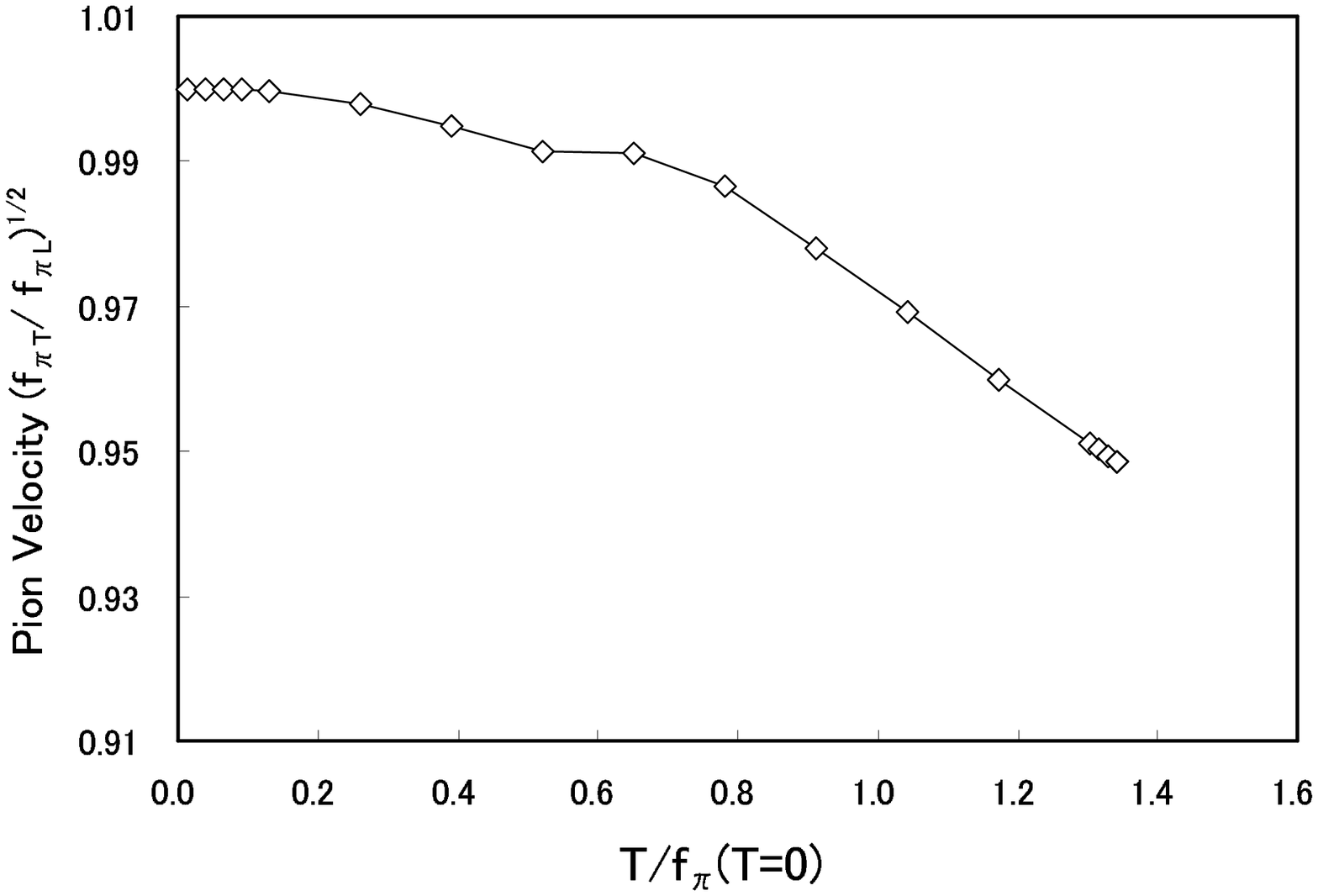}}
%\epsfxsize=0.61\textwidth
%\centerline{\epsfbox{velocity.eps}}
\caption{Temperature dependence of the pion velocity.}
\label{fig:5}
\end{figure}

The chiral condensate $<\bar\psi\psi>$ can be evaluated by the common manner with the zero temperature case.
The temperature dependence of the order parameters are
shown in Fig. \ref{fig:4}. 
$|<\bar\psi\psi>|^{1/3}/f_{\pi}(T=0)$ keeps almost constant value $2.8$ at low temperature 
$T/f_{\pi}(T=0)\leq 0.7$ and decreases rapidly at $T/f_{\pi}(T=0)\geq 1.2$. 
Finally, $|<\bar{\psi}\psi>|^{1/3}$ vanishes at $T=T_c$. 
%%%%%%%%%%%%%%%%%%%%%%%%%%%%%%%%%%%%%%05/5/23%%%%%%%%%%%%%%%%%%%%%%%%%%%%%%%%%%%%%
The transverse and longitudinal pion decay constants: $f_{\pi T}$ and $f_{\pi L}$
take almost same value at low temperature.
The difference between them is less than $1\%$ for $T/f_{\pi}(T=0)\leq 0.4$.
However, for high temperature over $T/f_{\pi}(T=0)\geq 0.8$, the difference of two decay
constants gradually becomes visible.
%%%%%%%%%%%%%%%%%%%%%%%%%%%%%%%%%%%%%%%%%%%%%%%%%%%%%%%%%%%%%%%%%%%%%%%%%%%%%
Here, we can  consider pion velocity in the
thermal bath. 
The pion momentum $\bar{p}$ and the pion energy $p_0$ satisfy the
dispersion relation:
\be
\label{eq:dispersion}
p_0^2=\F{Z_{\pi T}}{Z_{\pi L}}|\bar{p}|^2.
\ee
and pion's squared velocity $v^2$ is defined from the ratio of
pion wave function renormalization factors, or decay constants:
\be
\label{eq:vsquare}
v^2=Z_{\pi T}/Z_{\pi L}=f_{\pi T}/f_{\pi L}.
\ee
The temperature dependence of the pion velocity  $v$ is drawn in
Fig.\ref{fig:5}. It shows that pion travels at the velocity of light in the
low temperature region, and that it slows down slightly as temperature 
rises. 
At $T=T_c$, the pion velocity is slowed down to $v/c\approx 0.95$. 
 
\begin{figure}[htb]
\centerline{\includegraphics[width=9.15cm]{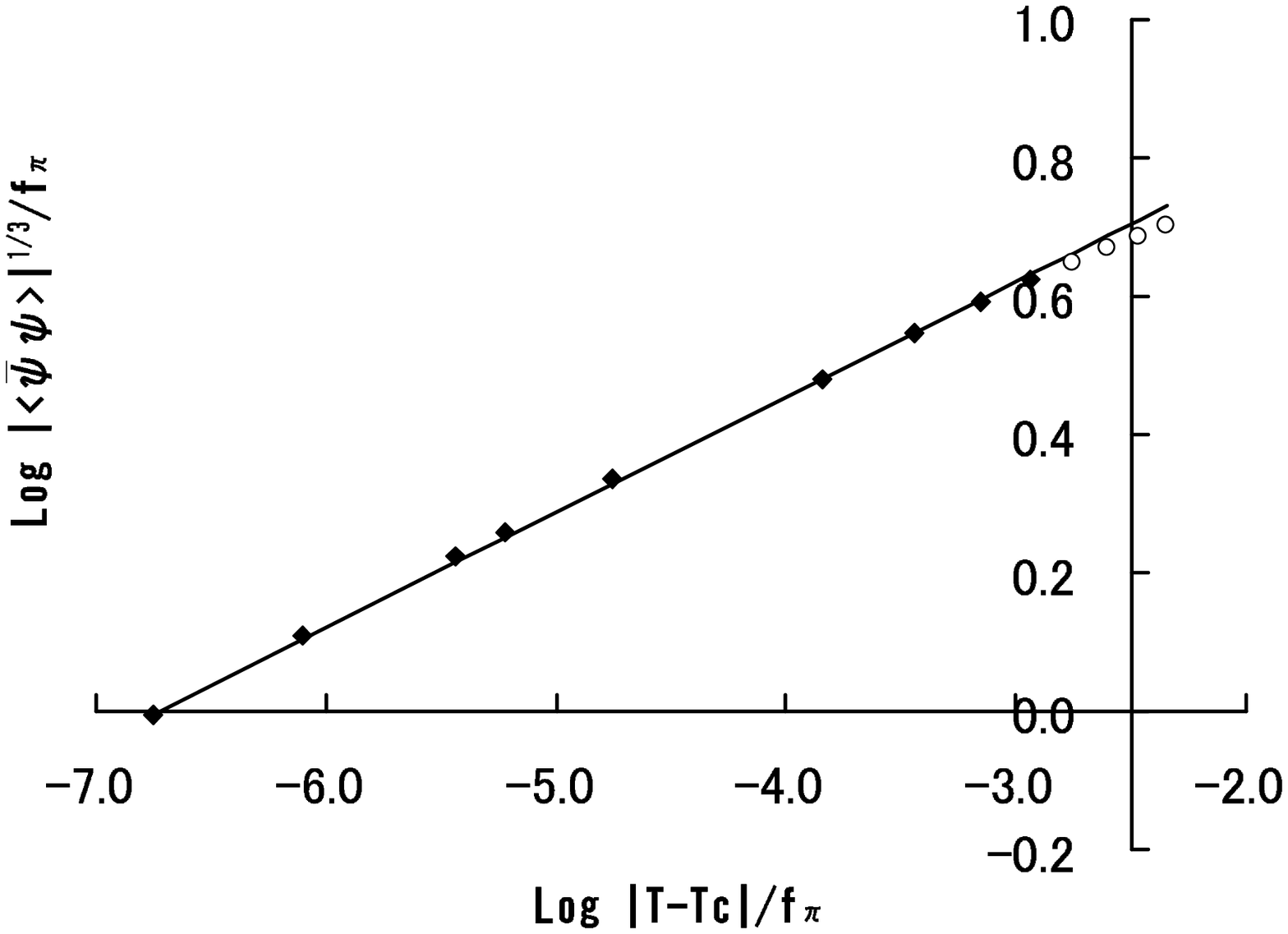}}
%\epsfxsize=0.61\textwidth
%\centerline{\epsfbox{log-3.eps}}
\caption{The relation of $\log |T-T_c|/f_{\pi}$ and $\log|<\bar\psi\psi>|^{1/3}/f_{\pi}$. 
The black points are the datum used in the linear regression, 
and the white ones are not used.}
\label{fig:6}
\end{figure}

Next, we want to know the critical exponents from the temperature dependence
of the order parameters.
From the construction in sec.\ref{sec:2}, our effective potential $V$ is invariant 
under the transformation:
$\bar\psi\psi\rightarrow -\bar\psi\psi, m_0\rightarrow -m_0, \sigma_0\rightarrow-\sigma_0$.
Thus, the chiral condensate in (\ref{eq:def_condensate}) is considered to behave as:
\be
<\bar\psi\psi>=\sigma_0f(\sigma^2_0(T)).
\ee
$<\bar\psi\psi>$ and $\sigma_0$ vanish at $T=T_c$, 
however, $f(0)$ keeps some finite value. 
Therefore, the critical exponent of the chiral condensate is derived by
\be
\label{eq:def_exponent}
|<\bar\psi\psi>(T)|\sim C\sigma_0(T)\sim |T-T_c|^\nu,
\ee
where $C$ is some constant. 
The critical exponents of $f_{\pi L}$ and $f_{\pi T}$ are estimated in the same way. 
$Z_{\pi L}$ and $Z_{\pi T}$ in (\ref{eq:fpiFT}) never go to zero when $T\rightarrow T_c$,
 so the exponents are determined by the temperature dependence of $\sigma_0(T)$, same as in 
 (\ref{eq:def_exponent}).
The critical exponents for
the chiral condensate $<\bar\psi\psi>$ and decay constants $f_{\pi
L}$ and $f_{\pi T}$ should be universal.
The logarithms of $|<\bar\psi\psi>|^{1/3}/f_{\pi}$ and $|T-T_c|/f_{\pi}$ are plotted in
Fig.\ref{fig:6}. 
We draw the line, $\log|<\bar\psi\psi>|=\nu \log|T_c-T|+C'$, where 
$\nu$ and $C'$ are determined by the ordinary least squares, 
using the black points in Fig.\ref{fig:6}. 
We can find the
critical exponent: $\nu=0.48$.

At zero temperature, it is verified that physical quantities do not depend too much on the choice of 
the IR regularization of the running coupling constant. \cite{rf:BSamp} 
\  
So we similarly need to examine the $\Lambda_1$ dependence of $T_c$ to 
guarantee the confidence of our results. In Ref.\citen{rf:Takagi},
$t_f=\ln (\Lambda_1/\Lambda_{\rm QCD})^2$ dependence of the numerical results for the QCD phase
structure is also discussed. It is shown that $T_c$ at $\mu=0$ are
stable for $t_f=0.6, 0.5$, and $0.4$. In our calculation we find
$T_c=120$ MeV, $124$ MeV, and $129$ MeV for $t_f=0.6, 0.5$, and $0.4$
respectively.

\subsection{Approximation scheme for the thermal mass}

So far, we carry out the calculations at finite temperature
introducing the gluon thermal mass in the HTL approximation. 
This is the most striking difference from many previous works in the SDEs.
There are several attempts to introduce the gluon thermal mass.
We finally carry out the calculation in other
approximations, and show how the difference is reflected to the result. 
In Table \ref{table:3}, we compare our result with other approximation schemes 
for the thermal screening effect. 

\begin{table}[htb]
\caption{The critical temperature and the critical exponent in several approximations.}
\label{table:3}
\begin{center}
\begin{tabular}{c|cc}
\hline 
\hline 
Scheme & $T_c$ (MeV) & $\nu$ \\
\hline
HTL approx.& $124.1$& $0.48$ \\
$|p_0|/|\bar p|$ leading & $98.6$& $0.47$\\
Static limit of HTL& $154.3$& $0.50$ \\
Neglected  & $171.9$& $0.50$ \\
\hline
\end{tabular}
\end{center}
\end{table}

The values `HTL approx' in Table \ref{table:3} are our results 
and the case labeled `$p_0/|\bar{p}|$ leading' is the
well known form of the Landau damping,\cite{rf:LeBellac} \ 
\be
\label{eq:p_0/pleading}
m_L^2=\F13g^2T^2\left(N_c+\F12N_f\right),\qquad
m_T^2=\F\pi 4m_L^2\F{|p_0|}{|\bar p|}.
\ee
This approximation has been applied to investigate the color superconductivity at high 
density.\cite{rf:colorsuper} \ Third one is the 
static limit of the HTL approximation,\cite{rf:SDEfiniteT1,rf:Takagi} \ i.e.
\be
m_L^2=\F13g^2T^2\left(N_c+\F12N_f\right),\qquad
m_T^2=0.
\ee
In the last case, the screening effect are dropped: 
\be
m_L^2=m_T^2=0.
\ee
In Ref.\citen{rf:SDEfiniteT2} and Ref.\citen{rf:SDEfiniteT3}, the chiral phase transition at finite temperature/density has discussed
in this approximation.
$T_c$ in the second case is the lowest in our results. 
It is supposed that the factor $|p_0|/|\bar{p}|$ in
(\ref{eq:p_0/pleading}) causes large $m_T$ for small $|\bar p|$ (or large 
$p_0$) and strong screening effects. 
The critical exponents in the third case and the last case are found to be $\nu=0.50$.
It seems that the critical exponent, at least in our approximation, depends on the treatment of  the magnetic mass.
The deviation from our result are at most $4\% $ for the critical exponent, 
while $20\sim 40\% $ for the critical temperature.  
The temperature dependences of the order parameters for each method are
shown below. In Fig.\ref{fig:7}, the temperature dependences of $|<\bar{\psi}\psi>|^{1/3}$ is plotted. 
It tells us that the difference by approximation for the thermal mass appears at $T/f_{\pi}(T=0)\geq 0.4$. 
In each case $|<\bar{\psi}\psi>|^{1/3}$  begins to decrease at $T/f_{\pi}(T=0)\approx T_c/f_{\pi}(T=0)-0.4$, and rapidly goes to zero
over $T/f_{\pi}(T=0)\approx T_c/f_{\pi}(T=0)-0.1$. 

\begin{figure}[htb]
\centerline{\includegraphics[width=9.2cm]{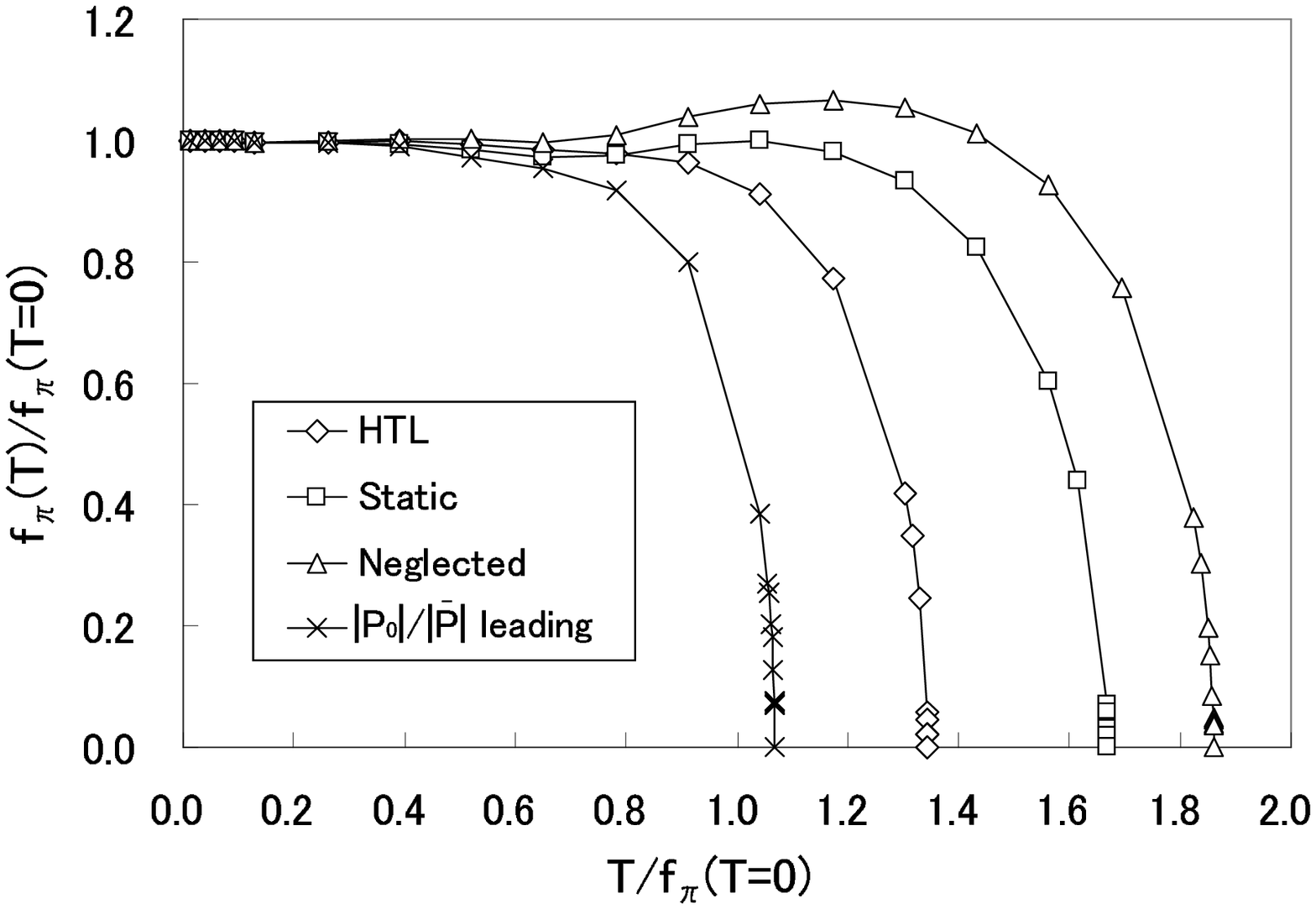}}
%\epsfxsize=0.61\textwidth
%\centerline{\epsfbox{piT.eps}}
\caption{Temperature dependence of $f_{\pi}(T)/f_{\pi}$ for each scheme.}
\label{fig:8}
\end{figure}

%%%%%%%%%%%%%%%%%%%%%%%%%%%%05/5/23%%%%%%%%%%%%%%%%%%%%%%%%%%%%%%%%%%%%%%%%%%%
In Fig.\ref{fig:8}, we plot $f_{\pi}(T)/f_{\pi}(T=0)$, 
where $f_{\pi}(T)$ is so-called `space-time averaged' decay constant,
\cite{rf:SDEfiniteT2} defined by $f_{\pi}(T)\equiv(f_{\pi
L}^2+3f_{\pi T}f_{\pi L})^{1/2}/2$.
It tells that the effects of the thermal mass appear at 
$T/f_{\pi}(T=0)\geq 0.4$. 
There is a definite peak in the `Neglected' case.
%In the peak $f_{\pi}(T)/f_{\pi}(T=0)$ is about $1.04$.
In SDEs, the temperature dependence of $f_{\pi}(T)$ is estimated, not that of $f_{\pi L}$ and
$f_{\pi T}$ one by one.\cite{rf:SDEfiniteT2} \   
In Ref.\citen{rf:SDEfiniteT2}, the dependence is calculated by
neglecting the thermal mass and there is a peak at
$T/f_{\pi}(T=0)\approx1.2$. 

\begin{figure}[htb]
\centerline{\includegraphics[width=9.2cm]{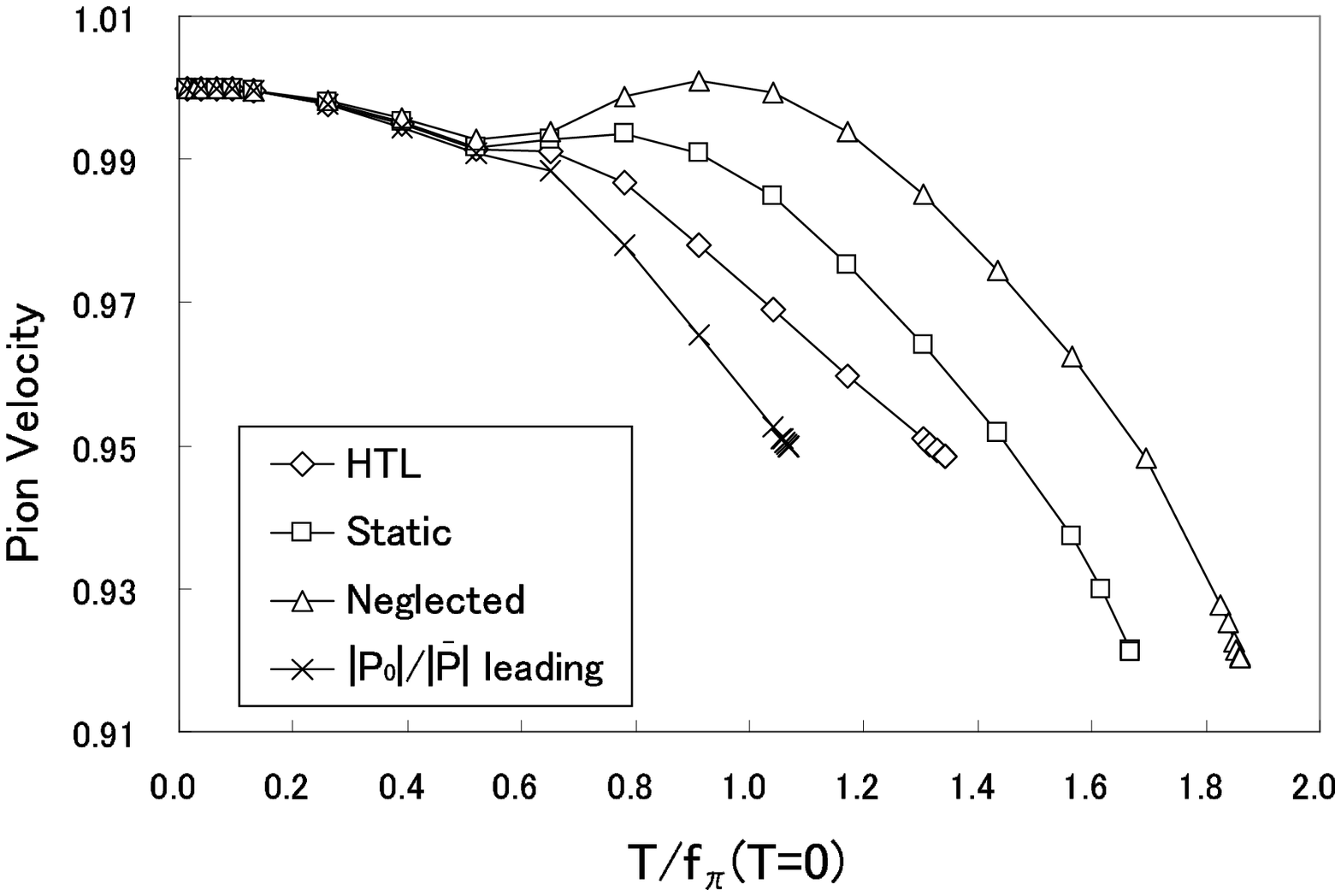}}
%\epsfxsize=0.61\textwidth
%\centerline{\epsfbox{piV.eps}}
\caption{Temperature dependence of pion velocity $v$ for each scheme.}
\label{fig:9}
\end{figure}

The quantitative difference becomes evidently for the pion velocity $v$. 
In Fig.\ref{fig:9},  the difference of how $v$ is slowed down is plotted. 
$v/c$ is finally reduced to $0.92, 0.92,0.95$, and $0.95$ for `Neglected', `Static limit', `HTL', 
and `$|p_0|/|\bar{p}|$ leading' respectively at each $T_c$. 
$v_c$ at $T=T_c$ hardly depend on $T_c$. It tells that $v/c$ at $T_c$
depends obviously whether magnetic mass is introduced or not.
There are peaks in the `Neglected' case and `Static limit' case,
and in the former case, $v/c$ is increased to about $1.00$ again near $T/f_{\pi}(T=0)=0.9$.
 
%%%%%%%%%%%%%%%%%%%%%%%%%%%%%%%%%%%%%%%%%%%%%%
\begin{figure}[htb]
\centerline{\includegraphics[width=9.2cm]{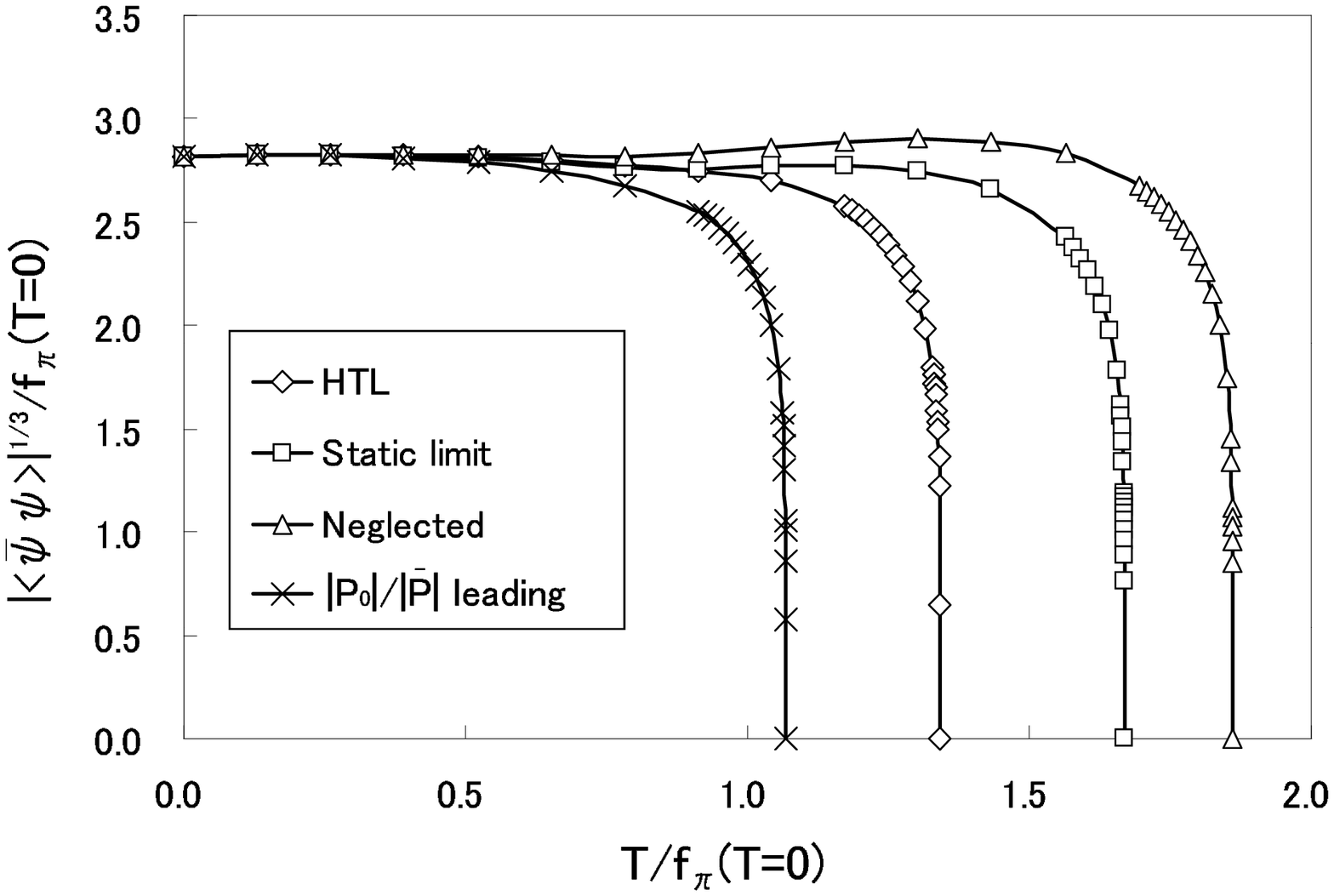}}
%\epsfxsize=0.61\textwidth
%\centerline{\epsfbox{condensate.eps}}
\caption{Temperature dependence of $|<\psi\bar{\psi}>|$ for each scheme.}
\label{fig:7}
\end{figure}

\section{Summary and Discussions}
\label{sec:4} 
In this paper we analyzed the chiral phase structure of finite temperature QCD for 3-flavors by
solving the sharp cutoff ERG flow equation. We derived the equation in the
ladder approximation.
The auxiliary fields $\sigma$ and $\pi$ were introduced at the scale $\LA$, 
which is larger than $\LQ$. 
It was shown that the dependence of $f_{\pi}$ on $\LA$ was considerably improved by the momentum scale expansion.
The gluon thermal mass is
incorporated by the HTL approximation. As a result the critical
temperature $T_c$ has become
about $124$ MeV, which is considerably lower value than the one calculated
by neglecting the thermal mass. We also make certain that $T_c$ and the
temperature dependence of the order parameters depend on
the approximation scheme for the thermal mass.
Our result shows that the screening effect is an important factor when
the chiral phase structure of finite temperature QCD is discussed.  

In this paper, we replaced the gauge sector with the momentum dependent
four fermi operator induced by one gluon exchange.
Of course, the gauge dynamics in finite temperature QCD can be
incorporated solely by the ERG. Then we have to employ the gauge 
invariant ERG\cite{rf:gaugekeep} or MSTI.\cite{rf:gaugeERG} \ 
As mentioned in \S\ref{sec:2}, both attempts will demand 
too much effort of human as well as machines.

Including the instanton effect is the one of the interesting extensions of this work.
There are various possible scenarios about the connection between $U(1)_A$ symmetry and
the chiral phase transition.\cite{rf:U1chiral} \ 
However, in the ERG framework $U(1)$ symmetry breaking can not be incorporated
straightforwardly
because of the IR cutoff.\cite{rf:U1A} \ 
To study the effect of the anomaly, we have to add the instanton induced multi fermi 
term to $\Gamma_{\Lambda}$ at the initial scale. 

The phase transition seems to be second order, or very weak first
order. We also studied the temperature dependence of the order
parameters for each approximation scheme for the thermal mass.
In the HTL, the chiral condensate and two $\pi$ decay constants $f_{\pi L}$ and $f_{\pi T}$
vanished for $T\approx124$ MeV $(T/f_{\pi}(T=0) \approx 1.34)$, that is the critical temperature.
When $T=0$, $f_{\pi L}$ and $f_{\pi T}$ have the same value. However, as 
temperature rises higher than $T/f_{\pi}(T=0)\approx 0.8$ and approaches $T_c$, the difference becomes
evident. The pion velocity $v$ is determined from the ratio of the decay
constants, and it decreases slightly as temperature approaches $T_c$.

%In other words, as shown in (\ref{eq:fpiFT}) and (\ref{eq:vsquare}), the temperature dependence of
%$f_{\pi L}$ and $f_{\pi T}$ is related to $v$ and $\sigma_0$.
%The physical reasons for the temperature dependence of these two quantities seem to be 
%different since $v$ never goes to zero even at $T=T_c$. Probably,  the rate of the decreasing of 
%$\sigma_0$ gets larger than that of $v$ near $T=90$ MeV, which may maximize $f_{\pi T}$. 

It is one of the attractive subjects to apply our method to the case of 
finite density. In the passed few years, it has been clarified that the vacuum of hot/dense QCD has rich structure,
e.g. color super conducting phase, color-flavor locking phase, and so
on.
\cite{rf:CSC} \ The application of our method to the finite density case 
seems to be straightforward.

Needless to say, HTL is based on perturbative QCD and it can not be said that this is truly proper method in
the temperature region for the chiral phase transition, at most $T\approx
100$ MeV. However, this is just our first step of the exploration of the QCD phase structure by ERG. 

\section*{Acknowledgments}
We would like to thank H. Kodama for the collaboration in the early
stage.

\appendix
\section{Thermal Threshold Functions}
\label{appendix:A}
In this appendix we give the explicit form of thermal threshold
functions  $I_{k(L,T)}$
in (\ref{eq:ERGZPILT})-(\ref{eq:ERGg2}).
First, we show what corresponds to the $O(p^0)$ component and  that of
the spatial components.

\bea
&&I_0=2N_fN_c\F{T}{\pi^2}{\sum_n}'\sqrt{\Lambda^2-\omega_n^2}
\F1{(\Lambda^2+m^2)},\\
&&I_{1T}=N_fN_c\F{\Lambda T}{\pi^2}{\sum_n}'\F{\omega_n^2+m^2}{(\Lambda^2+m^2)^2},\\  
%&&\Lambda\F d{d\Lambda}Z_{\pi L}=
%-2y^2\F{(\Lambda^2+m^2)\zeta^0-2\zeta_1}{(\Lambda^2+m^2)^3}\theta (\LA -\Lambda),\\
%\label{eq:ERGZT}
&&I_{2T}=
-\F23N_fN_c\F{\Lambda^2 T}{\pi^2}{\sum_n}'\sqrt{\Lambda^2-\omega_n^2}
\F{\Lambda^2+3m^2+2\omega_n^2}{(\Lambda^2+m^2)^3} %%\theta (\LA -\Lambda),
\eea
The temporal threshold functions are a little complicated because of
discretized expansion (\ref{eq:disexpansion1}) and
(\ref{eq:disexpansion2}).
$I_{1L}$ and $I_{2L}$ are given by
\bea
&&I_{1L}=N_fN_c\F{\Lambda T}{\pi^2}{\sum_n}'\sqrt{\Lambda^2-\omega_n^2}
\left(2\F{|\omega_n|}{(\Lambda^2+m^2)^2} -\F{J_1}{2\pi
T}\right) \nonumber\\
&&\qqqquad+N_fN_c\F\Lambda{\pi^3}
\sqrt{\Lambda^2-\pi^2T^2}K_1, \\
&&I_{2L}=N_fN_c\F{\Lambda^2 T}{\pi^2}{\sum_n}'\sqrt{\Lambda^2-\omega_n^2}
\left(-\F{|\omega_n|}{2\pi T(\Lambda^2+m^2)}J_1+\F1{4\pi^2T^2}J_2\right) \nonumber\\
&&\qqqquad-N_fN_c\F{\Lambda^2}{\pi^3}\sqrt{\Lambda^2-\pi^2T^2}\left(\F{\pi T}{(\Lambda^2+m^2)}K_1-\F1{2\pi T}K_2\right),\\
\eea
where $J_1$, $J_2$, $K_1$ and $K_2$ are defined as:
\bea
&&J_1\equiv\F3{\Lambda^2+m^2}-\F4{\Lambda^2+m^2+4\pi
T\omega_n+4\pi^2T^2}+\F1{\Lambda^2+m^2+8\pi
T\omega_n+16\pi^2T^2}, \\
&&J_2\equiv\F1{\Lambda^2+m^2}-\F1{\Lambda^2+m^2+4\pi
T\omega_n+4\pi^2T^2}+\F1{\Lambda^2+m^2+8\pi T\omega_n+16\pi^2T^2}, \\
&&K_1\equiv\F2{\Lambda^2+m^2}-\F1{\Lambda^2+m^2+8\pi^2T^2},\\
&&K_2\equiv-\F1{\Lambda^2+m^2}+\F1{\Lambda^2+m^2+8\pi^2T^2}.
\eea
At $T=0$, $I_{kL}$ and $I_{kT}$ reduce to an identical
form. They are given by 
\bea
&&I_0=\F{\Lambda^2}{2\pi^2}\F1{\Lambda^2+m^2},\\
&&I_1=\F{\Lambda^2}{3\pi^3}\F{\Lambda^2+3m^2}{(\Lambda^2+m^2)^2},\\
&&I_2=-\F{\Lambda^4}{4\pi^2}\F{\Lambda^2+2m^2}{(\Lambda^2+m^2)^3}.
\eea

%\section*{Acknowledgements}
%We would like to thank ...........

%\appendix
%\section{First Appendix} %Empty argument \section{} yields `Appendix'. 
%
%\section{Second Appendix}

\end{document}